\documentclass[9pt,sigconf,letterpaper]{acmart}
\usepackage[T1]{fontenc}

\usepackage[english]{babel}
\usepackage{graphicx}
\usepackage{booktabs}
\usepackage{xspace}
\usepackage[utf8]{inputenc}
\usepackage{wasysym}
\usepackage[ruled,vlined]{algorithm2e}
\usepackage{pdflscape}
\usepackage{enumitem}
\usepackage{verbatim}
\usepackage{verbatim}

\usepackage{subcaption}
\usepackage{balance}
\usepackage{float}
\usepackage{xcolor}
\usepackage{color}
\usepackage{colortbl}
\SetExtraKerning{encoding=*,font=*}{\textemdash={120,120}} 

\newcommand{\todo}[1]{}
\renewcommand{\todo}[1]{{\color{red} TODO: {#1}}}
\newcommand{\TODO}[1]{}
\renewcommand{\TODO}[1]{{\color{red} TODO: {#1}}}

\def \paragraph [#1] {\vspace{3pt}\noindent{\textbf{#1}\quad}}%

\newcommand{\tkc}[1]{}
\renewcommand{\tkc}[1]{{\color{red} <-CITE{#1}}}

\newcommand{\stanford}{Stanford\xspace}
\newcommand{\merit}{Merit\xspace}
\newcommand{\gn}{GreyNoise\xspace}
\newcommand{\orion}{Orion\xspace}

\title{Cloud Watching: Understanding Attacks Against Cloud-Hosted~Services}
\author[]{Liz Izhikevich}
\affiliation[]{Stanford University}
\author[]{Manda Tran}
\affiliation[]{Stanford University}
\author[]{Michalis Kallitsis}
\affiliation[]{Merit Network, Inc.}
\author[]{Aurore Fass}
\affiliation[]{Stanford University, CISPA Helmholtz Center for Information Security}
\author[]{Zakir Durumeric}
\affiliation[]{Stanford University}
\renewcommand{\shortauthors}{Liz Izhikevich, Manda Tran, Michalis Kallitsis, Aurore Fass, \& Zakir Durumeric}

\copyrightyear{2023}
\acmYear{2023}
\setcopyright{acmlicensed}\acmConference[IMC '23]{Proceedings of the 2023 ACM Internet Measurement Conference}{October 24--26, 2023}{Montreal, QC, Canada}
\acmBooktitle{Proceedings of the 2023 ACM Internet Measurement Conference (IMC '23), October 24--26, 2023, Montreal, QC, Canada}
\acmPrice{15.00}
\acmDOI{10.1145/3618257.3624818} \acmISBN{979-8-4007-0382-9/23/10}
\begin{document}

\ccsdesc[500]{Networks}
\ccsdesc[500]{Security and privacy~Network security}
\ccsdesc[300]{Security and privacy~Intrusion detection systems}
\keywords{cloud, security, honeypot, darknet, scanning}


\begin{abstract}
Cloud computing has dramatically changed service deployment patterns. In this work, we analyze how 
attackers identify and target cloud services in contrast to traditional enterprise networks and network telescopes. Using a diverse set of cloud honeypots in 5~providers and 23~countries as well as 2~educational networks and 1~network telescope, we analyze how IP address assignment, geography, network, and service-port selection, influence what services are targeted in the cloud.
We find that scanners that target cloud compute are selective: they avoid scanning networks without legitimate services and they discriminate between geographic regions. Further, attackers mine Internet-service search engines to find exploitable services and, in some cases, they avoid targeting IANA-assigned protocols, causing researchers to misclassify at least 15\% of traffic on select ports. Based on our results, we derive recommendations for researchers and operators.
\end{abstract}

\maketitle

\section{Introduction}
To understand attacker behavior, the networking and security communities have long analyzed the unsolicited traffic received by network telescopes (large swaths of unused IP addresses passively capturing unsolicited traffic)~\cite{singh2004automated, nawrocki2021far,kramer2015amppot, wustrow2010internet, mirai, durumeric2014internet, irwin2013source, chatziadam2014network, pang2017malicious, lagraa2017knowledge, 2004-moore-t20, prajapati2021shedding, cabana2021threat, akiyoshi2018detecting, malecot2013carna}. However, recent work has increasingly hinted that many of the conclusions about attacker behavior drawn from large network telescopes may not transfer to production networks where vulnerable services live in practice~\cite{richter2019scanning, soro2019darknets, irwin2013baseline, spoki, pham2011honeypot, pouget2005advantages}.

In this work, we investigate how attackers identify and exploit services in one largely unstudied type of network---\emph{cloud environments}---and how the malicious traffic seen by cloud hosts differs from that seen by Internet telescopes and education networks. Cloud environments  like Amazon~\cite{aws_ec2}, Google~\cite{google_ec2}, and Alibaba~\cite{alibaba_ec2} are notably different than other networks. First, cloud networks are dense: more than one third of publicly-exposed IPv4 services (around 100~million services) are hosted in a cloud environment~\cite{censys-2015}. Second, cloud providers host services from multiple owners with a range of security postures and business importance in a shared and recycled IP address space. Third, services in the cloud often follow non-traditional deployment patterns (e.g., many services live on non-IANA assigned ports~\cite{lzr}).

Using a set of interactive honeypots deployed by \gn across 5~cloud providers in 23~countries along with 1~network telescope and 2~education networks, we analyze how network type and provider, geography, service-port selection, and IP address assignment affect how services are scanned and exploited. Beyond differences in network type, we show that the cloud's recycled IP address space inadvertently impacts the security of cloud-hosted services. 
For example, attackers, including botnets, send orders of magnitude more (or less) traffic, depending upon a service's IP addresses' structure.  
Past ownership also affects observed behavior: IP addresses that previously hosted services that were indexed by Shodan~\cite{shodan} or Censys~\cite{censys-2015} attract a significantly different set of scanners and are targeted by 7~times more exploits than IP addresses that have never hosted search engine indexed services. 

The cloud's geographic diversity also influences the services that attackers target and the measurement conclusions drawn from a honeypot; attackers tailor usernames and passwords towards specific geographic regions, particularly in Asia Pacific.
Attackers who target cloud services often avoid scanning networks without legitimate services, creating a blind spot for telescopes to important attacker activity.
We repeat our analyses across two years of data and find that attacker preferences remain relatively stable over time. 
\looseness=-1





Without knowledge of confounding behaviors and statistical testing, researchers can easily misattribute differences in attacker behavior seen within cloud environments. Grounded in our analysis, we derive recommendations for both researchers and operators. For example, researchers should be wary of relying on only network telescopes for understanding network behavior and researchers should not directly compare traffic between individual honeypots, as attacker biases require statistical validation to extract larger trends. Operators should monitor unexpected ports/protocols, since attacker traffic may be unanticipated; and continue to monitor IP reputation, since scanners send an order of magnitude more traffic to IPs found on Shodan or Censys. We release our dataset of scanning traffic targeting the cloud to enable future research.

\section{Related Work}

A significant fraction of Internet measurement research use Internet telescopes, honeypots, and passive network analysis to understand topics that range from attacker behavior to Internet outages. While several prior studies have hinted that attackers exhibit bias during target selection~\cite{irwin2013baseline,spoki, pham2011honeypot, pouget2005advantages, soro2019darknets, moore2002network, richter2019scanning}, there has been little focus on cloud networks specifically. 
Our work builds off of existing research in three areas: telescope measurements, cloud computing, and Internet scanning, which we describe here.

\vspace{3pt}
\noindent
\textbf{Telescopes.} Network telescopes, also known as darknets, have been used to understand Internet background radiation~\cite{wustrow2010internet,pang2004characteristics}, malicious scanning patterns~\cite{irwin2013baseline,irwin2013source,chatziadam2014network,pang2017malicious,durumeric2014internet,anand2023aggressive}, DDoS attacks~\cite{nawrocki2021far,kramer2015amppot}, worms~\cite{singh2004automated}, and botnets~\cite{mirai,torabi2020inferring}. 
To ensure scientific validity, researchers have extensively studied the caveats of telescope deployment: understanding how the size~\cite{moore2002network}, network~\cite{soro2019darknets}, and geographic location~\cite{irwin2013baseline,spoki,pham2011honeypot,pouget2005advantages,soro2019darknets} of darknets influence unsolicited scans and attacks. Calibration studies have primarily compared darknets to other darknets~\cite{soro2019darknets, gadhia2015comparative,nishijima2021verification,irwin2013baseline,irwin2015observed} or darknets to honeypots within similar networks~\cite{francois2008activity, akiyoshi2018detecting}. However, our work shows that attackers targeting the cloud frequently avoid darknets altogether and exhibit unique preferences within cloud networks.

Most closely related, in 2019, Richter et~al.\ showed that there are significant differences between scans that target darknets and a CDN~\cite{richter2019scanning}. 
Griffioen et~al., investigated amplification DoS attacks and found little overlap in amplification DoS attacks between the cloud and a darknet~\cite{griffioen2021scan}. 
Our work also shows that telescopes which do not collect payloads, mistakenly assume that scans only target IANA-assigned protocols. Further, we demonstrate that telescopes that collect payloads but reside in networks that do not emulate real services (e.g.,~\cite{spoki}) are avoided by scanners.  

\vspace{3pt}
\noindent
\textbf{Honeypots in the Cloud.}
Several recent studies have measured Internet activity using cloud-hosted honeypots. 
For example, Kelly et al.,  Bove et al., and Brown et al. study network differences amongst honeypot traffic, but only within the US~\cite{kelly2021comparative,bove2019} or when aggregating different geographic regions across different networks~\cite{brown2012HoneypotsIT}. 
We find that there are several surprising confounds that affect the traffic that a cloud honeypot receives, and that without statistically rigorous hypothesis testing, it is easy to draw incorrect conclusions. For example, our work shows that some reported prior results, such as network preferences~\cite{brown2012HoneypotsIT}, are not statistically significant. Most prior works~\cite{kelly2021comparative,brown2012HoneypotsIT,bove2019} do not perform statistical tests in their analysis, making it unclear to what extent their observed differences are statistically significant or due to chance, and how their results can direct future work.


\vspace{3pt}
\noindent
\textbf{Internet Scanning.}
While prior work~\cite{durumeric2014internet,richter2019scanning,spoki,blaise2020detection,wustrow2010internet} has shown that the vast majority of Internet scanners target a small sub-sample of the IPv4 address space, to the best of our knowledge, no work has investigated how attackers target towards specific service histories within the cloud. 
Most closely related, Irwin~\cite{irwin2013baseline} finds scanners targeting port~445 are less likely to scan broadcast addresses in telescope networks and Moura et~al.~\cite{moura2014bad} finds neighboring IP in ISPs are more likely to engage in spam and phishing attacks. Similarly,  Bodenheim~\cite{bodenheim2014impact} measure the impact of the Shodan service search engine on IoT devices and Raftopoulos et al.~\cite{raftopoulos2015dangerous} show that Internet scanning can lead to compromised hosts. 

The variety of scanning traffic targeting neighboring services requires statistically-rigorous comparisons.
Francois et~al.~\cite{francois2008activity} propose a parametric method for detecting significant changes in telescope networks using a threshold that must be manually determined for each distribution type. Instead, we present a non-parametric method tailored towards small sample sizes, which cloud vantage points often provide. 
Last, our study is motivated in part by recent findings on real-world service deployment. Bano et~al.~\cite{bano2018scanning} noted that protocols oftentimes run on unassigned ports. Izhikevich et~al.~\cite{lzr} found that services on unassigned ports are more likely to be vulnerable. We are the first work that analyzes how attackers scan unexpected services. We show that prior studies that only rely on popular honeypot frameworks~\cite{cowrie, tpot, kippo} or non-reactive telescopes---and therefore assume that scans are targeting the IANA-assigned protocol---miss at least 15\% of scan traffic.


\section{Methodology}
\label{sec:methodology}

\begin{table*}[h!]
\footnotesize
\centering
\begin{tabular}{lllll|rr}
\toprule

Network & \# Geo & Geographic Region & Vantage Points & Collection & \# Unique & \# Unique \\ 
 & Regions & Country (State) Codes & (IPs) per Region &  Method & Scan IPs & Scan ASes \\ 
\midrule
Hurricane Electric & 1 &
US (OH)
& 256 & \gn 
& 130,103 & 8,278 \\
AWS & 16 &
US (OR), US (CA), US (GA), BR, BH,
& 4 or 2~(HTTP) & \gn 
& 99,566 & 7,142 \\
& & FR, IE, DE, CA, AU, SG, IN, KR, JP, & & \\
& & HK, ZA & & \\
Azure & 3 & 
US (TX), SG, IN
&4 or 2~(HTTP) & \gn 
& 19,893 & 2,477 \\
Google & 21 & 
US (NV), US (UT), US (CA), US (OR),
& 4 or 2~(HTTP) & \gn 
& 103,067 & 7,481 \\
& & US (VA), US (SC), US (IA), QC, CH,  & & \\
& & NL, DE, GB, BE, FI, AU, ID, SG, KR, & &  \\
& & JP, HK, TW & & \\

Linode & 7 &
US (CA), US (NY), UK, DE, IN, AU, SG
& 4 or 2~(HTTP) & \gn 
& 72,235 & 5,984 \\
\midrule
\stanford & 1 & US (West) & 64 & Honeytrap 
& 105,045 & 6,177 \\
AWS & 1 & US (West) &  64 & Honeytrap 
& 99,445 & 7,603 \\
Google & 1 & US (West) & 64 & Honeytrap 
& 93,119 & 7,947 \\
\merit & 1 & US (East) & 64 & Honeytrap 
& 106,988 & 6,315 \\
Google & 1 & US (East) & 2 & Honeytrap 
& 18,064 & 1,802\\
\midrule
\orion & 1& US (East) & 475K &Telescope & 5,147,050 & 24,835 \\  


\bottomrule
\end{tabular}
\vspace{3pt}
\caption{\textbf{Vantage points}---%
\textnormal{We analyze scanning traffic targeting 8 unique networks (5~cloud, 2~education, and 1~telescope), spanning 23 countries across North America, Europe, and Asia Pacific. We use three different scanning traffic collection methods described in Section~\ref{sub:sec:data}.
We report the number of unique IPs and ASes that scan each vantage point between July 1--7, 2021. 
}}
\vspace{-10pt}
\label{table:data_breakdown}
\end{table*}

To understand the differences in network attacks seen within cloud environments, we analyze traffic seen by honeypots in different networks, geographic regions, and with different service configurations. In this section,
we describe our primary data sources, how we differentiate benign from malicious scanning traffic, how we minimize the risk of harm during our experiments, the statistical tests we use to compare scanning traffic, and how we validate the temporal stability of our results.

\subsection{Vantage Points}
\label{sub:sec:data}

To obtain a diverse set of vantage points, we use existing honeypots from \gn, deploy our own honeypots, and use the \orion Network Telescope.
We summarize all our data sources, including their geographic location and size, in Table~\ref{table:data_breakdown}. 
We publicly release our data at:
\url{https://scans.io/study/cloud_watching}.

\vspace{3pt}
\noindent
\textbf{\gn honeypots.}
\gn deploys honeypots across multiple cloud providers and geographic regions. 
\gn honeypots are assigned public IPv4 addresses, which are not publicly known.\footnote{The data we release contains honeypot IP addresses that are no longer in use.}.
\gn uses Cowrie~\cite{cowrie}, an interactive honeypot, to collect SSH (ports 22, 2222) and Telnet (23, 2323) attempted login credentials 
For all other ports, \gn completes the TCP or TLS handshake and records only the first received payload.  
Each \gn honeypot hosts public vulnerable-looking protocol-assigned services on at least seven popular ports. 

\gn deploys a variable number of honeypots across different regions and protocols. 
To maximize the number of honeypots per region while also maintaining consistency, we limit our analysis to regions that contain at least 4~SSH honeypots, 4~Telnet honeypots, and 2~honeypots for all other protocols (non-SSH and non-Telnet protocols nearly always only have 2 honeypots per region). We filter to include only geographic regions and networks that collect data in both 2020 and 2021 for cross-validation purposes. After filtering, there remain honeypots across 5~cloud providers---AWS, Google, Azure, Linode, and Hurricane Electric---and 23~countries across North America, Europe, and Asia Pacific. 


\vspace{3pt}
\noindent
\textbf{Honeytrap honeypots.}
To understand how attackers that target clouds also target other networks with legitimate services, we use two existing /26 IPv4 networks of honeypots at two educational institutions: \stanford and \merit. The honeypots use the Honeytrap~\cite{honeytrap} framework for traffic collection and configure it to collect the first UDP payload or the first TCP payload after completing a TCP handshake. To eliminate biases when directly comparing the education and cloud honeypots, we deploy an additional 64~IPv4 Honeytrap honeypots in a Google geographic region located near \stanford,  64~IPv4 honeypots in an AWS geographic region located near \stanford, and 2~IPv4 honeypots in a Google geographic region near \merit. We do not compare traffic between \gn and Honeytrap honeypots given their different software configurations.

\vspace{3pt}
\noindent
\textbf{\orion network telescope.}
To understand how attackers that target clouds also target other networks without legitimate services, we analyze scanning activity targeting a network telescope. 
Network telescopes/darknets typically do not host any services, receive traffic on all ports and IP addresses, and only record the first packet of a connection (i.e., they do not complete the TCP layer~4 handshake).
To compare the scanning activity of a telescope with scanning activity targeting networks that host real services (e.g., educational networks or cloud providers), we use the \orion Network Telescope, which spans 475K IPv4 addresses (i.e., 1,856 /24 networks).
\looseness=-1

\vspace{3pt}
\noindent We discuss limitations of our vantage points in Section~\ref{sec:limitations}.

\label{sec:ethics}
\vspace{3pt}
\noindent
\textbf{Ethics.}
To minimize harm when deploying honeypots, we configure the honeypots to not expose services that are historically prone to being abused for amplification attacks (e.g., DNS open resolver).
Furthermore, our honeypots do not respond to UDP messages, ensuring that no UDP-based DDoS amplification attacks occur. 
The honeypots are also configured to be low-interaction, thereby limiting the size of responses and minimizing the chances of arbitrary code execution triggering a harmful zero-day amplification attack. 



\subsection{Identifying Malicious Traffic}
\label{sub:sec:suricata}


Not all network scanning is malicious. Multiple motivations exist behind unsolicited network scans: organizations collecting datasets~\cite{censys-2015, shodan}, academic groups conducting research~\cite{durumeric2013zmap} or performing vulnerability notifications~\cite{durumeric2014matter,li2016you}, malicious actors performing reconnaissance with the intent of later exploitation~\cite{le2009mitibox}, or malicious actors actively exploiting a service~\cite{mirai}. Understanding the true intent behind a network scan is challenging: \gn's mission is to identify scanning actors, yet 78\% of the scanning IPs that \gn  encountered in 2022 were classified as ``unknown''~\cite{greynoise_viz}.  

When possible, we classify whether a scan is malicious based on whether the scan attempts to (1) login or bypass authentication, or (2) alter the state of the service (e.g., run a shell command). Our definition does not account for reconnaissance scanning that may have delayed malicious intent. Throughout our analysis, we refer to ``scanners'' as those for whom the scanning intent is unknown (e.g., any scanner that targets a telescope that does not collect payloads) and ``attackers'' as those for whom malicious intent has been verified (e.g., a scanner that sends a malicious payload). While an attacker is also a scanner, we make the distinction to maintain precision in our claims. While detecting malicious behavior is easy for protocols that request authentication (e.g., SSH, Telnet), non-authentication based protocols (e.g., HTTP) pose a challenge. For example, while the HTTP protocol is commonly used for sending benign \texttt{GET} requests~\cite{spoki}, many exploits are also delivered over HTTP, including the critical Log4Shell (CVE-2021-44228) vulnerability~\cite{log4shellHTTP}. 

To detect malicious payloads that attempt to bypass authority or alter the state of a service for non-authentication-based protocols, we use Suricata~\cite{suricata}, an open-source network intrusion detection system providing 32K detection rules.
Following Suricata documentation recommendations \cite{suricataRules}, we manually filter for rules that limit false positives (e.g., rules that do not rely on a static set of blocklisted IPs or ports).
To eliminate false positives, we (1) manually inspect the subset of rules that trigger alerts on payloads and (2) only keep rules that are triggered when the corresponding payload is verified as bypassing authority or altering the state of service. 
Our final rule set belongs in the following Suricata class types: trojan-activity, web-application-attack, protocol-command-decode, attempted-user, attempted-admin, attempted-recon, bad-unknown, misc-activity. 
The Suricata rules used are found on Pastebin~\cite{suricataTriggeredRules, suricataTriggeredRulesREADME}.

Suricata labels 6\% (10.2K) of distinct HTTP payloads in our dataset as malicious.
Overall, we identify that 34\% of traffic does not attempt to bypass authentication when targeting 23/Telnet, 24\% does not bypass authentication when targeting 22/SSH, and 75\% of payloads do not send exploits to HTTP/80. Thus, prior works~\cite{torabi2020inferring,lagraa2017knowledge,irwin2013source} whose methodology assumes that all traffic destined towards commonly vulnerable ports (e.g., Telnet/23) is malicious, and all traffic destined towards commonly benign ports (e.g., HTTP/80) is benign, likely misclassify at least a quarter of traffic. 

\subsection{Comparing Vantage Points}
\label{sub:sec:chi_meth}

As we will show in the next section, there are confounding biases when differentiating neighboring targets, making the use of statistical tests necessary when comparing attacker activity across vantage points. To find significant differences between the traffic that targets different honeypots, we perform the non-parametric chi-squared statistical test~\cite{pearson1900x}. 

To identify statistically significant differences, we use a p-value of 0.05 and apply Bonferroni correction to accommodate the comparisons across all vantage points. Often, Bonferroni correction shrinks p-values by several orders of magnitude. Since the p-value is only a measure of statistical certainty, we use Cramér's V~\cite{cramer1946contribution} to calculate the effect size (denoted by~$\phi$), which indicates the strength of statistical difference: the larger the effect size, the more different the distributions. The magnitudes of effect sizes do not have predefined limits (e.g., not all $\phi < 0.3$ represents a small effect). Rather, magnitudes are derived using the chi-statistic and the degrees of freedom within the chi-test, both of which depends upon the number of unique values being compared. Thus, identical $\phi$ values can represent different effect sizes if the degrees of freedom between two tests are different. To promote understanding, for each test we report the effect size alongside its magnitude. 

The chi-square test expects a minimal number of variables with an expected frequency of zero, so that it does not inaccurately mark distributions as significantly different due to a small skew in the long-tail of near-zero frequencies.
As there is a long tail of scanning actors (e.g., on average, the top~3 ASes that send the most traffic of all 680 ASes account for 37\% of all traffic sent to each \gn honeypot), we limit the degrees of freedom and ensure the expected frequency of a variable is larger than zero (an important requirement for chi-squared tests). Concretely, we always choose the most popular 3 values for each characteristic (e.g., top~3 payloads, top~3 scanning ASes) for each vantage point and perform the chi-squared test on the union of all unique top~3 characteristics across vantage~points.
Studying the top~3 values decrease bias towards small distributional differences.\footnote{The long-tail of ASes/payloads that scan each honeypot restricts the number of top popular values we can compare at at time.  For example, while the top-3 ASes account for 37\% of all scanning traffic, the top-5 account for 42\% and the top-100 account for 70\%. Thus, expanding evaluation to even the top-5 ASes increases the number of near-zero frequency variables by over 200\%, significantly increasing bias towards small distributional-differences; studying top-3 decreases bias.}

Our analysis includes many dimensions of comparisons. To simplify, we focus on 3 popular assigned protocols: Telnet (the most popular protocol used by botnets~\cite{mirai}), HTTP and SSH (the two protocols responsible for over 90\% of ASCII payloads sent by network scanners~\cite{spoki}). 
We also consider the possibility of scanner behavior varying across non-IANA assigned ports, and  report HTTP results\footnote{We only analyze HTTP across all ports, since malicious HTTP packets can be fingerprinted without needing application-layer specific interaction across all ports.} independent of port number (i.e., ``HTTP/All Ports'').

Across vantage points, we use the chi-squared test to compare scanning traffic using the following axes: \emph{who} (i.e., which ASes are scanning), \emph{what} (i.e., what are the top usernames/passwords/payloads being attempted), and \emph{why} (i.e., the maliciousness of traffic).
When comparing who is scanning, we often identify scanning actors by their autonomous system, as opposed to IP address, to account for scanning campaigns that rely on multiple source IP addresses (e.g., Censys~\cite{censysScanningOptOut}).
When comparing payloads, we directly compare usernames and passwords for SSH and Telnet, and directly compare the full payload after removing ephemeral values (i.e., Date, Host, and Content-Length fields) for HTTP. 


\subsection{Temporal Stability}
\label{sub:sec:meth_temporal}
We compare scanning traffic across all three sources of vantage points (cloud, educational, and network telescopes) using data collected during the first week of July~2021.
To verify that our results are consistent across time, we repeat our experiments using data from the first week of July~2020 or July~2022 (depending on the availability of vantage points at that time) and provide the results in Appendix~\ref{app:temporal}. Across the 3 years, the IP addresses of our honeypots remain consistent, while those of the \gn honeypots change. We supplement the results throughout the paper with a discussion on temporal similarities and differences.  

\section{Impact of IP Address Assignment}
\label{sec:sources}

Services hosted in the cloud live in a randomly-assigned and recycled IP space.
Cloud services acquire neighbors with a range of security postures, and they occupy IP addresses that have previously housed services with a range of reputations. 
In this section, we explore if and how a service's IP address and history influence what scanners target. We find that, indeed, attackers target neighboring\footnote{We define neighboring services to be services that reside in the same geographic region and network (i.e., from the same cloud provider, educational network, or network telescope), but do not necessarily share contiguously neighboring IP addresses.} 
identical services differently, such as sending a varying number of malicious payloads, usernames, and passwords. We explore what factors influence the services scanners target, and find that scanners predict network structures to filter for targets and mine Internet-service search engines to find exploitable services.

\subsection{Variation Across Neighboring Hosts}
\label{sub:sec:min_bias}
\begin{table}[t]
\centering
\resizebox{\columnwidth}{!}{%
\begin{tabular}{lrrrr}
\toprule

&
\multicolumn{2}{c}{SSH/22}
&
\multicolumn{2}{c}{Telnet/23}  \\ \cmidrule(r){2-3}\cmidrule(l){4-5}

Traffic & \% Neighborhoods &  Avg. $\phi$
 
 &\% Neighborhoods  &   Avg. $\phi$  \\
 
Characteristic & w/ dif distributions & &
 w/ dif distributions  &  \\
 & (n = 53) & & (n = 53)&  \\
	\midrule 
Top~3 AS	& 44\% & \cellcolor{red!10}0.31   & 38\% & \cellcolor{red!10}0.43   \\
Fraction Malicious	& 36\% & \cellcolor{blue!10}0.12  & 15\% &  \cellcolor{blue!10}0.12 \\
Top~3 Username	& 55\% & \cellcolor{yellow!10}0.22 &21\% &\cellcolor{red!10}0.24\\
Top~3 Password	& 4\%& \cellcolor{yellow!10}0.13 &19\% &\cellcolor{red!10}0.39  \\

\bottomrule
\end{tabular}
}
\resizebox{\columnwidth}{!}{%
\begin{tabular}{lrrrr}
\toprule

&
\multicolumn{2}{c}{HTTP/80}  
&
\multicolumn{2}{c}{HTTP/All Ports} 
\\ \cmidrule(r){2-3}\cmidrule(l){4-5}

Traffic &\% Neighborhoods  &   Avg. $\phi$
 & \% Neighborhoods  &  Avg. $\phi$  \\
Characteristic   & w/ dif distributions & 
  &  w/ dif distributions &    \\
   & (n = 61) & &  (n = 61)&  \\
	\midrule 
Top~3 AS	& 31\% & \cellcolor{red!10}0.43  & 42\% & \cellcolor{yellow!10}0.23 \\
Fraction Malicious	& 0\% & - & 19\% & \cellcolor{blue!10}0.04    \\
Top~3 Payloads	& 15\% & \cellcolor{red!10}0.39& 77\% & \cellcolor{yellow!10}0.17  \\
\bottomrule
\end{tabular}
}
\vspace{3pt}
\caption{\textbf{Attackers target neighboring services differently}---%
\textnormal{A different set of ASes attack neighboring services with different payloads, usernames, and passwords.  
We compare distributions using the chi-square methodology from Section~\ref{sub:sec:chi_meth} and color the effect sizes with the relative magnitude (i.e., blue=``small'', yellow=``medium'', red=``large'').}
}
\vspace{-10pt}
\label{table:ipaddr_effects}
\end{table}

Neighboring services in the cloud are scanned and attacked by a significantly different group of scanners and payloads. 
In Table~\ref{table:ipaddr_effects}, we compute the percentage of neighborhoods in the clouds that receive significantly different traffic using data from \gn vantage points for the following traffic characteristics: the top 3 ASes that send traffic (malicious or not), the fraction of malicious traffic, the top 3 usernames and password attempts for SSH and Telnet, and the top 3 payloads across all traffic for HTTP.
A significantly different set of ASes target neighboring services (large effect size, $\phi$=0.43). 
For example, one of four identical services in the Linode network Singapore geographic region is targeted by three orders of magnitude more unique scanning IPs from Axtel Networks (ASN~\,6503) compared to the other services (large $\phi$= 0.82).
Thousands of scanner IP addresses belonging to the Tsunami botnet~\cite{tsunamiBot} only target a single IP address in the Hurricane Electric /24 honeypot network. 

Across neighboring services, attackers attempt different payloads when bypassing authentication of services, including different usernames (e.g., large $\phi$=0.24 targeting Telnet/23) and different passwords (e.g., large $\phi$=0.39 targeting Telnet/23). 
For example, attackers send an order of magnitude more payloads that attempt an HTTP POST user login request to only one of four identical honeypot services in the Azure network Singapore geographic region (large $\phi$=0.61).
In the next sections, we explore two reasons that contribute to significant differences amongst neighboring services: IP address structure and  Internet service search engines.

\subsection{IP Address Structure}
\label{sub:sec:impacts_target_scan}
\label{sec:ip_addrRisk}

Service operators and attackers treat IP addresses differently. 
While service operators often assign IP addresses to hosts at random (e.g.,  dynamic host configuration, cloud-assigned virtual machine addresses), scanners and attackers use the IP address to predict the presence of targets.
We identify which IP address structures scanners are most likely to target in the cloud by (1) using the network telescope to identify scanning patterns (given its substantially larger sample size) and (2) validating the existence of the same pattern in the cloud. 

Scanners avoid IP addresses that are believed to not host services in both the telescope and cloud. 
We compare the number of scanners across neighboring IP addresses in the telescope, which we plot in Appendix~\ref{app:filt_net}.
We observe that scanners are 3.5~times less likely to target an IP address structure that is likely reserved for broadcasting purposes (i.e.,  ending in a ``.255'') compared to other IP addresses, on seven of the top ten most consistently targeted ports.
Scanners targeting port~445 in the cloud also exhibit a similar bias: scanners are between 1.2~(Google) to 3.5~times (Linode) less likely to target a ``.255'' IP address.
However, unlike the telescope, we find no significant evidence of ``.255'' avoidance on other ports in the cloud, perhaps due to the different set of attackers that target clouds and telescopes (Section~\ref{sub:sec:net}).


In the telescope, scanners that avoid broadcast-type addresses for one octet are equally likely to avoid an IP address with other ``255'' octets (e.g., \texttt{x.A.255.0/24}). 
The avoidance is significant: for example, scanners targeting 7574/Oracle are 61~times less likely to target an IP with a ``255'' octet; and 9~times less likely for 445/SMB. We hypothesize that incorrect filtering of broadcast addresses, in which the position of the ``255'' octet is not checked, may be responsible for the observed preference.
Since none of our cloud honeypots have IP addresses with a ``255'' octet that does not appear at the end, we leave to future work to validate the existence of this pattern in the cloud. 
\looseness=-1

Botnets exhibit less intuitive, yet still significant, preferences in both the telescope and cloud. For example, when targeting port~22 in the telescope, the Mirai botnet and scanners from the bullet hosting provider PonyNet (ASN\,53667) are one order of magnitude more likely to choose the first address of a /16 (e.g., \texttt{x.B.0.0}) as its first scanning target compared to any other address. 
Within our  Hurricane Electric /24 honeypot network, the Tsunami botnet~\cite{tsunamiBot} is one order of magnitude more likely to target a single IP address. 
Thus, random IP address assignment leaves some services unknowingly more vulnerable to botnet attacks than others. 


\subsection{Internet Service Search Engines}
\label{sec:ip_add_rep}
The recycled address space of the cloud assigns services to IPs that previously hosted unrelated services.
In this section, we investigate how attackers use the most-frequently scanning Internet service search engines~\cite{li2020survey}---Censys~\cite{censys-2015} and Shodan~\cite{shodan}---to find services. 
We discover that attackers are more likely to scan and exploit IPs previously indexed by Internet-service search engines.

\vspace{3pt}
\noindent
\textbf{Methodology.} \quad
To measure if attackers use Internet service search engines, we deploy additional Honeytrap~\cite{honeytrap} honeypots emulating SSH/22, Telnet/23, and HTTP/80 services across the following three groups of IPs:
\vspace{-.17cm}
\begin{itemize}[noitemsep, leftmargin=*]
    \setlength\itemsep{.1em}
    \item \textbf{Control group honeypots} are deployed on 8~IPs that have not had services in at least 4~years. We block Censys and Shodan from accessing the Honeytrap services for the duration of the experiment by blocklisting the IPs they scan with. 
    \item \textbf{Previously leaked honeypots} are deployed on 7~``recycled'' IPs that have hosted an HTTP/80 scanning information page for at least two years (while conducting Internet-wide scans). While Censys and Shodan previously advertised the HTTP/80 service on these hosts, we block Censys and Shodan from accessing the Honeytrap services for the duration of the experiment. 
    \item \textbf{Leaked honeypots} are deployed on 18~IPs that have not had services in at least 4~years. At the beginning of our experiment, we systematically leak the Honeytrap services: we split the 18~IP addresses in groups of 3 IPs and allow either Censys or Shodan to find only one of the three emulated services: SSH/22, Telnet/23, or HTTP/80. For example, one group of 3~IP addresses only allows Censys to discover their HTTP/80 service, one group only allows Censys to discover their SSH/22 service, 
    one group only allows Shodan to discover their SSH/22 service, etc. 
\end{itemize}

\vspace{-3pt}
\noindent
By systematically ``leaking'' services to the two most popular Internet service search engines~\cite{most_pops_search}, we test how search-engines influence the services that attackers target. 
When comparing and presenting our results, we exclude scanning traffic from Censys and Shodan so that increases in scanning traffic are not due to the Censys/Shodan scanners themselves.
To perform our experiment, we do not deploy honeypots in the cloud because our experiment requires un-tainted service histories, and we do not control the service history of cloud IPs.
Thus, we deploy the honeypots in a network we control: \stanford. 
While this network is not a cloud network, our results in Section~\ref{sub:sec:net} show that scanners that target the cloud are similar to scanners that target education networks---roughly 89\% of IPs that target the cloud also target the education network. There is no significant difference in the payloads or fraction of malicious traffic. Thus, our analysis of scanners targeting the \stanford network can likely be extrapolated to also characterize scanners that target the cloud. 


\begin{table}[t]
\centering
\footnotesize
\begin{tabular}{lllll}
\toprule

 Service &Traffic & Censys & Shodan & Previously\\
& & Leaked & Leaked & Leaked \\
\cmidrule(r){3-5}
& & \multicolumn{3}{c}{Fold Increase in Traffic per Hour}  \\

\midrule
HTTP/80 & All & \textbf{7.7*} & 15.7* & 17.2* \\
& Malicious & \textbf{4.0*} & 5.8 & \textbf{7.3} \\
\midrule
SSH/22 & All & 2.4 & \textbf{2.6*}  & \textbf{1.5*}  \\
& Malicious & 2.5 & \textbf{2.8*} & \textbf{1.7*} \\
\midrule
Telnet/23 & All & 72.6*  & \textbf{1.06*}  & 201 \\ 
& Malicious & \textbf{1.6*} &  \textbf{1.3*} & \textbf{1.8} \\


\bottomrule \end{tabular}
\vspace{3pt}
\caption{\textbf{Impact of Internet Service search engines}---
	\textnormal{Attackers are more likely to attack a service that is currently, or has been previously, indexed by Censys or Shodan. Statistically significant increases are marked in bold and traffic distributions that are significantly different from our control group's traffic distribution (e.g., exhibit spikes of---but not necessarily overall---increased volume) are indicated by \texttt{*}.}
	}
\label{table:searchEngines}
\vspace{-15pt}
\end{table}

\vspace{3pt}
\noindent
\textbf{Attackers use Internet-service search engines.} \quad
We observe two primary attacker behaviors that target leaked services.
First, across protocols, scanners and attackers are significantly\footnote{We use a one-sided Mann-Whitney U test to evaluate whether the volume of traffic per hour that targets leaked services is stochastically greater than the volume targeting the control group. We only discuss significant results.} 
more likely to target a service that is currently, or has been previously, leaked (Table~\ref{table:searchEngines}).
For example, HTTP/80 services listed on Censys or Shodan are attacked with 7.3~times more malicious traffic per hour compared to non-leaked services. SSH/22 services leaked on Shodan are attacked with 2.8~times more malicious traffic per hour than non-leaked services, and 1.6~times more for Telnet/23 services found on Censys. 

Second, we observe that attackers are significantly\footnote{We use the Kolmogrov-Smirnov test to compare the distributions of the average volume of traffic per hour targeting leaked and non-leaked services. Upon manual verification, we determine that the spikes of traffic are the underlying cause of the difference in distributions.} 
more likely to increase the number of ``spikes'' of traffic towards leaked services. 
In other words, scanners and attackers are more likely to only briefly scan a leaked service, likely after it has been found by the attacker on a search engine. 
For example, scanners send significantly more spikes of traffic towards Shodan-leaked HTTP/80 and Censys-leaked Telnet/23 services compared to non-leaked services. 
Spikes of traffic often carry unique brute force logins; attackers will attempt on average 3~times more unique SSH passwords on leaked compared to non-leaked services.


A different set of ASes target leaked HTTP/80 services.
For example, while three ASes---Avast (ASN\,198605), M247 (ASN\,9009), and CDN77 (ASN\,60068)---conduct nmap~\cite{nmap} scans against our non-Censys-leaked HTTP/80 honeypots, they actively \textit{avoid} all Censys-leaked HTTP/80 honeypots.  
Interestingly, the nmap scanners also target the previously leaked honeypots, implying that the nmap scanners source only up-to-date information from Censys. 
We do not find significant differences in the ASes that scan leaked and non-leaked SSH/22 and Telnet/23 services, nor do we find significant differences in the most popular payloads targeting leaked SSH/22 and Telnet/23 services. 

Attackers targeting a specific set of protocols also exhibit search-engine preferences (Table~\ref{table:searchEngines}): attackers targeting HTTP/80 rely more on Censys (4.0~times increase in traffic per hour) while attackers targeting SSH/22 rely more heavily on Shodan (2.8~times increase in traffic per hour). Attackers targeting Telnet/23 use both Censys and Shodan (1.3--1.6~times increase in traffic per hour) but rely on search engines less than attackers targeting SSH and HTTP\@.

\subsection{Discussion and Summary}

The vulnerability of services in the cloud are dependent on their randomly-assigned IP address due to differences in attacker proclivities. 
Scanners guess network structures, botnets latch on to individual targets, and malicious actors rely on Censys and Shodan to identify targets to brute-force attack.
Consequently, neighboring services see significant differences in malicious payloads. Hence, researchers who deploy honeypots in the cloud can also inadvertently observe dramatically different patterns in attacker behavior.

\vspace{3pt}
\noindent
\textbf{Temporal consistency.}\quad
Over the years, scanners and attackers have consistently exhibited preferences between neighboring targets.
In 2013, Irwin~\cite{irwin2013baseline} found that scanners targeting port~445 were less likely to scan broadcast addresses in telescope networks, which we confirm is still the case.
When analyzing our data from 2020, we observe the same patterns as in 2021 (e.g., scanners and attackers still originate from different ASes and send different payloads towards neighboring services), which we detail in Appendix~\ref{app:sub:sec:temporal_neighboring}.


\vspace{3pt}
\noindent
\textbf{Filtering attacker preferences.}\quad
\label{sub:sec:scanning_bias}
In the rest of our analysis, we account for attacker preferences for certain IPs and network structures by (1) using multiple honeypots in each region and (2) comparing the median expected values (e.g., the median number of packets sent by an AS within a group of honeypots) across groups. We elect not to compare the intersection of all scanning events within a group of honeypots, since the majority of scanning campaigns conduct sub-sampled Internet-wide scans and are not expected to target all honeypots within a region~\cite{durumeric2014internet,richter2019scanning,spoki,blaise2020detection}. 


\section{Geographies and Providers}
\label{sec:net_geo}

\begin{table}[t!]
\centering
\resizebox{\columnwidth}{!}{%
\begin{tabular}{ll|lr|lr|lr}
\toprule
&
&
\multicolumn{2}{c}{AWS}
&
\multicolumn{2}{c}{Google} 
&
\multicolumn{2}{c}{Linode}

\\ \cmidrule(r){3-4}\cmidrule(l){5-6}\cmidrule(r){7-8}

Traffic & Protocol & Most Dif.&  Avg. $\phi$  &   Most Dif. &  Avg. $\phi$   & Most Dif.&  Avg. $\phi$ \\
 & &   Region  &     &  Region &   &   Region  &    \\
	\midrule 
Top~3  & SSH/22 &  AP-JP & \cellcolor{red!10}0.68 & AP-SG   &\cellcolor{yellow!10}0.16 & AP-SG &\cellcolor{yellow!10} 0.27 \\
AS & TEL/23 & AP-AU &  \cellcolor{red!10}0.50  & - & -  & - & -  \\
& HTTP/80 &  AP-IND & \cellcolor{red!10}0.53   & AP-ID  & \cellcolor{red!10}0.47   & - & - \\
& HTTP/All & AP-SG & \cellcolor{yellow!10} 0.21  & AP-AU & \cellcolor{yellow!10} 0.23  & US-CAL & \cellcolor{yellow!10} 0.28  \\
\midrule
Top~3 	& SSH/22 & AP-JP  & \cellcolor{yellow!10} 0.47 & - & -  & - & - \\
Username & TEL/23& AP-AU  & \cellcolor{red!10}0.56 & - & -  & - & - \\
\midrule
Top~3 Password & TEL/23& CA-TOR & \cellcolor{red!10}0.52  & - & -  & AP-SG & \cellcolor{red!10}0.50 \\
\midrule
Top~3 
 & HTTP/80 & AP-HK & \cellcolor{yellow!10} 0.31 & AP-ID  & \cellcolor{yellow!10} 0.27 & AP-SG & \cellcolor{yellow!10} 0.35  \\
Payload  & HTTP/All& AP-HK& \cellcolor{yellow!10} 0.32 & AP-ID & \cellcolor{yellow!10} 0.25 & AP-ND &\cellcolor{red!10} 0.47 \\
\midrule
Fraction  & SSH/22 & AP-AU & \cellcolor{blue!10} 0.13 & - & - & - & - \\
Malicious & TEL/23& AP-AU & \cellcolor{blue!10} 0.16  & - & -  &  - & - \\
& Any/All& - &-& AP-JP & \cellcolor{blue!10} 0.04 & - & - \\

\bottomrule \end{tabular}
}
\vspace{3pt}
\caption{\textbf{Geographic regions with most different traffic patterns}---
\textnormal{When comparing all geographic regions against each other, Asia Pacific (AP) regions exhibit the largest statistically significant deviations of traffic distributions compared to other geographic regions within the same network. We mark the absence of  statistically significant results with a ``-''. We color the effect sizes with its the relative magnitude (i.e., blue=``small'', yellow=``medium'', red=``large''). As discussed in Section~\ref{sub:sec:chi_meth}, identical $\phi$ values can have different effect sizes given the degrees of freedom per experiment.} 
}
\vspace{-15pt}
\label{table:geo_effects}
\end{table}

Deploying services across multiple geographic regions and providers is remarkably simple in the cloud. 
In this section, we explore how attackers target services across different geographies and networks, after accounting for the biases that scanners exhibit when targeting neighboring services.
We find that attackers exhibit significant biases when scanning across continents or within Asia Pacific.
However, attackers rarely discriminate amongst different cloud networks within the same geographic region. 
Further, scanners and attackers that target the cloud are likely to avoid scanning networks that are publicly known to not host services (i.e., telescopes). 

\subsection{Discriminating Geographic Regions}
\label{sub:sec:geo_bias}
\label{sec:location}
\begin{table*}[t]
\centering
\footnotesize
\resizebox{\textwidth}{!}{%
\begin{tabular}{lllllllll}
\toprule

&
\multicolumn{4}{c}{SSH/22}
&
\multicolumn{4}{c}{Telnet/23}  \\ \cmidrule(r){2-5}\cmidrule(l){6-9}

Traffic  & \multicolumn{4}{c}{\% Similar Pairs of Regions in Same Geo-Region/Network} & \multicolumn{4}{c}{\% Similar Pairs of Regions in Same Geo-Region/Network}  \\
\cmidrule(r){2-5}\cmidrule(r){6-9} 
Characteristic & US (n=31) &  EU (n=19)  & APAC (n=40)  &Intercontinental (n=267)  &  US (n=31) &  EU (n=19)  & APAC (n=40)  &Intercontinental (n=267)    \\
	\midrule 
Top~3 AS	& 94\% & 100\% & 63\% & 70\% &100\% &100\% & 73\% & 81\%  \\
Frac Malicious	&  94\%& 100\%& 88\% & 83\% &100\%&100\% & 98\% & 99\%  \\
Top~3 Username	& 94\% & 100\% & 88\%  & 79\% & 100\% & 89\% & 75\% & 76\%  \\
Top~3 Password	& 100\%&100\%& 100\% &100\% & 100\% & 89\% & 73\% & 75\%   \\

\bottomrule \end{tabular}
}
\quad 
\resizebox{\textwidth}{!}{%
\begin{tabular}{lllllllll}
\toprule

&
\multicolumn{4}{c}{HTTP/80}  
&
\multicolumn{4}{c}{HTTP/All Ports} 
\\ \cmidrule(r){2-5}\cmidrule(l){6-9}
Traffic  & \multicolumn{4}{c}{\% Similar Pairs of Regions in Same Geo-Region/Network} & \multicolumn{4}{c}{\% Similar Pairs of Regions in Same Geo-Region/Network}  \\
\cmidrule(r){2-5}\cmidrule(r){6-9} 
Characteristic & US (n=31) &  EU (n=19)  & APAC (n=40)  &Intercontinental (n=267)  &  US (n=31) &  EU (n=19)  & APAC (n=40)  &Intercontinental (n=267)   \\
 
	\midrule 
Top~3 AS	& 97\% & 100\% & 85\% & 92\% & 91\% & 84\% & 44\% &39\%   \\
Frac Malicious &100\%& 100\%& 100\%&  100\% &100\%& 100\%& 100\%&  99\%  \\
Top~3 Payloads	& 94\% &100\% &90\% & 94\% & 50\% & 53\% & 20\% & 11\%  \\

\bottomrule \end{tabular}
}
\vspace{3pt}
	\caption{\textbf{Traffic similarities within and between geo-locations}---%
	\textnormal{Scanners targeting assigned services in regions within the US or EU nearly always originate from the same top~3 ASes and attempt the same most common payloads. However, geographic regions within Asia Pacific are much more likely to exhibit statistically significant variation in traffic characteristics.}
 }
\label{table:geo_sim}
\vspace{-10pt}
\end{table*}

We investigate how attackers consider geography when identifying targets in the cloud.
Attackers exhibit significant biases across continents and across the Asia Pacific region. However, contrary to prior work's inferences~\cite{bove2019} and telescope results~\cite{spoki}, they do not send significantly more or less malicious payloads within the US or EU.  
\looseness=-1

\vspace{3pt}
\noindent
\textbf{Methodology.}\quad
We compare traffic distributions from the \gn honeypots across geographic regions using the statistical methodology described in Section~\ref{sub:sec:chi_meth}.
We group continental regions in the same manner that AWS and Google group datacenters (i.e., North America, Europe, Asia Pacific).
We exclude Azure and Hurricane Electric due to their lack of geographic diversity in our dataset. 

\vspace{3pt}
\noindent
\textbf{Attackers discriminate among Asia Pacific.}\quad
Scanners and attackers exhibit the most significant preferences when targeting Asia Pacific. 
In Table~\ref{table:geo_effects}, we show that, across Asia Pacific, attackers attempt significantly different payloads than in other regions (large $\phi$ 0.27--0.47), including different usernames (large $\phi$ 0.47--0.56) and different passwords (large $\phi$ 0.50--0.52). 
For example, the top attempted Telnet usernames for most geographic regions are ``root'', ``admin'', and ``support.'' However, honeypots within the AWS Australia region see an order of magnitude less of those usernames, and are most targeted with ``mother'' and ``e8ehome,'' a credential often used by the Mirai botnet targeting Huawei devices~\cite{e8home}. 

There are also biases within the Asia Pacific region. Across Asia Pacific, scanners and attackers isolate specific sub-regions to avoid or target. For example, Emirates Internet (ASN\,5384) sends HTTP/80 post requests only towards honeypots located in Mumbai, India---the location closest to the United Arab Emirates in our dataset---while scans from SATNET (ASN\,14522) Ecuador target all geographic regions except for Mumbai. 

Attacker preferences are widespread throughout the Asia Pacific: 80\% of Asia Pacific region pairs are targeted with different distributions of HTTP payloads across all ports.
Scanners target significantly different regions of the Asia Pacific across all cloud providers: AWS, Google, Linode.
Attackers attempt significantly more different SSH and Telnet usernames between Asia Pacific geographic regions (large $\phi$ 0.47--0.56) than amongst neighboring services (Section~\ref{sec:ip_addrRisk}, large $\phi$ 0.22--0.24). 
However, when comparing top attempted passwords, fraction of malicious traffic, and scanning ASes,
scanners and attackers exhibit a similar magnitude of biases when targeting neighboring and inter-continental services. 

We do not find any consistent AS-geographic patterns that directly explain why Asia Pacific biases exist. 
For example, while attackers are \textit{less} likely to send malicious traffic in the Asia Pacific Azure and AWS regions (small $\phi < 0.16$), they are \textit{more} likely to send malicious traffic in Google's Asia Pacific region (small $\phi = 0.04$). 
Grouping too many autonomous cultures/governments (i.e., compared to grouping states and countries within North America) within the Asia Pacific---a common methodology in technology, politics, and commerce~\cite{apac}---might contribute to the variation.

\vspace{3pt}
\noindent
\textbf{Attackers do not discriminate between sub-regions within the U.S. and Europe.}\quad
Scanners exhibit significantly less biases when scanning within the US and EU (Table~\ref{table:geo_effects}). 
For example, the same set of ASes consistently target regions within the US or EU, and attackers do not send significantly more (or less) malicious payloads to a particular region.
While scanners send different payloads across 50\% of US and 53\% of EU geographic regions (Table~\ref{table:geo_sim}), the effect size is always smaller when compared to differences between Asia Pacific sub-regions. 
We observe scanners send an increased amount of Telnet payloads to the AWS Paris region, and more Android emulator commands to the AWS Frankfurt region. 
We find no significant differences in the median scanning traffic volume within or across continents. 
Our results are consistent with Section~\ref{sub:sec:net}, in which education networks located on the opposite coasts of the US see no significant differences in traffic. 



\subsection{Discriminating Network Types}
\label{sub:sec:net}

While attackers discriminate between and amongst certain geographic regions, they are unlikely to discriminate amongst different cloud providers in the same geographic region.\footnote{Due to lack of sufficient honeypots in different providers and regions within Asia Pacific, we are only able to verify this result in North America and Europe.}
However, we do find that many attackers that target networks that do have services (i.e., cloud, education) do not scan networks that are publicly known to not have services (i.e., network telescopes). 
Thus, consistent with prior results, we emphasize that researchers that rely on only telescopes are blind to an important scanning population that only targets and attacks real Internet services.

\vspace{3pt}
\noindent
\textbf{Methodology.}\quad
We compare traffic across networks using the methodology from Section~\ref{sub:sec:chi_meth}. 
To perform cloud-to-cloud comparisons, we use \gn data and compare only cloud honeypots that are located in the same city or state to minimize geographic biases (Table~\ref{table:net_vantages}). To avoid comparing data from different honeypot frameworks, we use the Honeytrap honeypots we deployed in AWS and Google geographically near the Honeytrap honeypots in the EDU networks to compare cloud and EDU networks. We use the Honeytrap honeypots in \stanford and \merit for the EDU--EDU comparison. When comparing education networks and the network telescope, we ensure that all honeypots are located in the US (which Section~\ref{sub:sec:geo_bias} shows minimizes bias). 

\begin{table}[t]
\centering
\footnotesize
\begin{tabular}{lcccc}
\toprule

\multicolumn{1}{c}{City} & \multicolumn{4}{c}{Cloud}  \\
\cmidrule(l){1-1} \cmidrule(l){2-5} 

 & AWS & Google & Linode & Azure \\ 
 \midrule
CA, US & $+$  &$+$  &$+$  & \\
GA, US & $+$  & & $+$  & \\
OR, US & $+$  & $+$  & & \\
TX, US & & & $+$  & $+$  \\
VG, US & & $+$  & & $+$  \\
FRA, GE & $+$  &$+$  &$+$  & \\

\bottomrule
\end{tabular}
\vspace{3pt}
\caption{
\textbf{Honeypots in multiple clouds}---%
\textnormal{When comparing scanner activity between networks, we only compare traffic destined towards vantage points located in the same city or state, in order to minimize geographic biases.}
}
\vspace{-15pt}
\label{table:net_vantages}
\end{table}
\begin{table}[t]
\centering
\footnotesize
\resizebox{\columnwidth}{!}{%
\begin{tabular}{ll|rr|rr|r}
\toprule
&
&
\multicolumn{2}{c}{Cloud--}
&
\multicolumn{2}{c}{Cloud--} 
&
EDU--\\

&
&
\multicolumn{2}{c}{Cloud}
&
\multicolumn{2}{c}{EDU} 
&
EDU\\
\cmidrule(r){3-4}\cmidrule(l){5-6}\cmidrule(r){7-7}

Traffic & Protocol & \# dif. region &  Avg. $\phi$  &   \# dif. region &  Avg. $\phi$   & \# dif. region \\
 
 & &  (n=10) & & (n=4) & & (n=1)  \\
	\midrule 
Top~3  & SSH/22 & 2& \cellcolor{blue!10} 0.11 & 3 & \cellcolor{red!10} 0.48 & 0   \\
AS & TEL/23 & 5& \cellcolor{blue!10} 0.21  & 0 & - &0   \\
& HTTP/80 & 3  & \cellcolor{blue!10} 0.15 &1 &\cellcolor{blue!10} 0.16 &0  \\
& HTTP/All & 6 & \cellcolor{blue!10} 0.21 & 2 & \cellcolor{blue!10} 0.10 & 0    \\
\midrule
Top~3 	& SSH/22 & 2 &\cellcolor{blue!10} 0.06  &$\times$&$\times$&$\times$ \\
User & TEL/23& 2 & \cellcolor{blue!10} 0.05 &$\times$&$\times$&$\times$   \\
\midrule
Top~3 & TEL/23& 0 & -  &$\times$&$\times$&$\times$ \\
Pwd &  SSH/22 & 0 & -  &$\times$&$\times$&$\times$  \\
\midrule
Top~3 
 & HTTP/80 & 4  & \cellcolor{blue!10}  0.19  & 1&\cellcolor{blue!10} 0.15 &0   \\
Payload  & HTTP/All& 6 &\cellcolor{blue!10} 0.23 &1 &\cellcolor{blue!10} 0.06 & 0  \\
\midrule
Frac  & SSH/22 &   1 & \cellcolor{blue!10} 0.01  &$\times$&$\times$&$\times$  \\
Mal & TEL/23& 2 & \cellcolor{blue!10} 0.02  &$\times$&$\times$&$\times$ \\
&  HTTP/80 & 0  &- & 0 &- & 0  \\
& HTTP/All& 0 & -& 0& -& 0  \\

\bottomrule \end{tabular}
}
\vspace{3pt}
\caption{\textbf{Differences across network types}---
\textnormal{Scanners that target cloud networks are unlikely to prefer a specific cloud (e.g., AWS versus Google), but are more likely to partially avoid education networks. Fields that cannot be calculated due to lack of payload collection are denoted by an $\times$. Effect sizes ($\phi$) are colored with their relative magnitude (i.e., blue=``small'', red=``large'').}
}
\vspace{-15pt}
\label{table:net_effects}
\end{table}

\vspace{3pt}
\noindent
\textbf{Scanners do not discriminate between networks with real services.}\quad
Although scanners significantly avoid the telescope network, we demonstrate in Table~\ref{table:net_effects} that scanners targeting assigned services within different cloud networks nearly always originate from the same top~3 ASes (small $\phi$<0.21) and attempt the same most common usernames and passwords (small $\phi$<0.06). We never see scanning ASes entirely ignore specific cloud regions. Zero cloud honeypots see a difference between the most popular SSH and Telnet passwords within a European or North American region. 
However, the majority of scanners that target \textit{unassigned} services (i.e., aggregating  across all ports and protocols)  originate from different ASes and attempt different payloads (small $\phi$=0.23). Nevertheless, the differences are much smaller than those seen across neighboring services (Section~\ref{sec:ip_addrRisk}) and those alluded to in prior work studying network telescopes~\cite{soro2019darknets}.
\looseness=-1

We never observe scanners significantly discriminating between education networks, even though the networks are located on opposite coasts of the US.
This shows that attacker discrimination of the telescope network is not geography-induced.
Scanners also do not significantly discriminate between cloud and education network: scanners always attempt the same usernames, passwords, payloads (small $\phi$ < 0.15), and send the same amount of malicious traffic. 

There is one exception. In 2021, scanners targeting SSH/22 in clouds were more likely to originate from different ASes than those that targeted education networks (large $\phi$=0.48). Six times more scanners from Chinanet (ASN\,4134) targeted the SSH/22 service in our education networks compared to cloud networks, while seven times more scanners from Cogent networks (ASN\,174) target the SSH/22 service in our cloud networks compared to our education networks. However, in 2022, we no longer saw significant difference between the scanners targeting SSH/22 in the cloud and education networks (Appendix~\ref{app:sub:sec:temporal_net}).
The absence of a difference implies that either (1) targeted-SSH events are an anomaly, or (2) targeted-SSH events ``spike'' (a pattern defined in Section~\ref{sec:ip_add_rep}) and are less likely to appear across all slices of time. 
The popular presence of SSH/22 in the clouds (e.g., AWS EC2 instances often come pre-configured with SSH/22) might contribute to attracting scanners in spikes.

\begin{table}[t]
\centering
\footnotesize
\begin{tabular}{lccc}
\toprule

Port & $\mid$ Tel $\cap$ Cloud $\mid$ & $\mid$ Tel $\cap$ EDU $\mid$ & $\mid$ Cloud $\cap$ EDU $\mid$ \\
 \cmidrule(l){2-2}\cmidrule(l){3-3}\cmidrule(l){4-4}
&  $\mid$ Cloud $\mid$ & $\mid$ EDU $\mid$ & $\mid$ Cloud $\mid$\\

\midrule
23 & 91\% & 96\% & 88\% \\
2323 & 53\% & 94\% & 83\%\\
80 & 73\%  & 86\% & 82\%\\
8080  & 80\%& 85\% & 90\%\\
21  & 29\%& 82\% & 94\%\\
2222  & 9\%& 82\% & 94\%\\
25 & 19\%& 79\% & 84\%\\
7547& 33\%  &71\%  & 97\%\\
22  &13\%& 60\% & 94\%\\
443 &30\%&44\%  & 81\%\\

\bottomrule
\end{tabular}
\vspace{3pt}
\caption{
\textbf{Scanners avoid telescopes}---%
\textnormal{Scanners that target the majority of popular ports at least once across any of our 440~cloud vantage points avoid scanning any of the 475K IPs in the telescope on the same port. However, the vast majority of scanners that target the cloud also target EDU networks.}
}
\vspace{-15pt}
\label{table:dark_cov}
\end{table}

\begin{table}[t]
\centering
\footnotesize
\begin{tabular}{lcc}
\toprule

Port & $\mid$ Tel-IPs $\cap$ Mal. Cloud-IPs $\mid$ &   $\mid$ Tel-IPs $\cap$ Mal. EDU-IPs $\mid$\\
 \cmidrule(l){2-2}\cmidrule(l){3-3}
&  $\mid$ Mal. Cloud-IPs $\mid$ & $\mid$ Mal. EDU-IPs $\mid$ \\

\midrule 
23 & 94\% & $\times$\\ 
2323 & 88\% & $\times$\\ 
80 & 84\% & 96\%\\  
8080  & 84\% & 97\% \\
2222  &3.6\% & $\times$\\
22  &7.5\% & $\times$\\
\bottomrule
\end{tabular}
\vspace{3pt}
\caption{
\textbf{Attackers targeting SSH-assigned ports in the cloud avoid telescopes}---%
\textnormal{A maximum of 7.5\% of attacker IPs that target SSH assigned ports at least once across any of our 440~cloud vantage points also scan any of the 475K IPs in the telescope on the same port. 
The majority of attacker IPs that target the education honeypots also target the telescope. 
Not every field can be calculated due to the manner in which payloads are or are not collected (Section~\ref{sub:sec:data}), denoted by an $\times$.
We do not perform the analysis between cloud and education networks, due to the small sample size of malicious scans that target the set of cloud honeypots that are located in the same geographic region as the EDU honeypots.}
}
\vspace{-10pt}
\label{table:mal_dark_cov}
\end{table}



\begin{table}[t]
\centering
\footnotesize
\begin{tabular}{ll|rr|rr}
\toprule
&
&
\multicolumn{2}{c}{Telescope--EDU}
&
\multicolumn{2}{c}{Telescope--Cloud}

\\ \cmidrule(r){3-4}\cmidrule(l){5-6}

Traffic & Protocol & \# dif. region &  Avg. $\phi$ & \#  dif. region &  Avg. $\phi$ \\
 
 & &  (n=2) & &  (n=3)  \\
	\midrule 
Top~3  & SSH/22 &  2  & \cellcolor{red!10}0.41  & 3 & \cellcolor{red!10}0.71  \\
AS & TEL/23 & 2& \cellcolor{red!10} 0.68 &3 & \cellcolor{red!10}0.82   \\
& HTTP/80  & 0& -& 2& \cellcolor{red!10}0.40   \\
& HTTP/All & 2 & \cellcolor{blue!10} 0.20  &3  &\cellcolor{blue!10} 0.30    \\

\bottomrule \end{tabular}
\vspace{3pt}
	\caption{\textbf{Different scanners target telescopes}---
	\textnormal{A significantly different set of ASes target telescopes, compared to clouds and education networks.  
	We color the relative magnitude (blue=``small'', red=``large'') of all effect sizes~($\phi$).}
	}
 \vspace{-10pt}
\label{table:tel_net_effects}
\end{table}

\vspace{3pt}
\noindent
\textbf{Scanners and attackers avoid telescopes.}\quad
Across the majority of popular ports, scanners that target networks with real services (i.e., clouds and education networks) are not seen in the network telescope.
In Table~\ref{table:dark_cov}, we compute the fraction of overlap between the IP addresses that target at least one cloud or education honeypot and the telescope;
only 13\% of IPs that target port~22 on any of our cloud honeypots send at least one packet to port~22 in the telescope. Only 44\% of scanners that target port~443 in one of our education honeypots also scan port~443 in the telescope. 
Scanners that target services hosted in education networks are more likely to target the telescope than those that target services in cloud networks (e.g., 71\% vs.\ 33\% on port~7547). 
We hypothesize this is due to \merit and \orion being located in the same autonomous system. 
Telnet/23 is the only service targeted by scanners that, for the most part, does not discriminate against telescopes: at least 91\% of the IPs that scan clouds and educational networks also scan the telescope. We hypothesize the lack of network preference is due to the prevalence of botnet scanning activity, which historically has not avoided unused IP address space~\cite{mirai,tsunamiBot}. 

Attackers targeting SSH-assigned ports also avoid telescopes.
In Table~\ref{table:mal_dark_cov}, we perform a similar analysis, but filter for scanners that send malicious payloads to cloud or education networks. 
Less than 10\% of attackers that target SSH-assigned ports on the cloud also target the telescope. 
A significantly different set of ASes scan telescopes; e.g., in Table~\ref{table:tel_net_effects}, ASes targeting Telnet/23 in telescopes and clouds differ with a large effect size of 0.82.
ASes geo-located in China actively avoid scanning the telescope; 12~times more unique scanners from China Mobile (ASN\,56046) and 2.5~times more unique scanners from Chinanet (ASN\,4134) target SSH/22 in our cloud and education honeypots compared to the telescope. 




Researchers studying honeypots located in cloud and education networks, as opposed to network telescopes, are more likely to encounter attackers targeting real services. 
In Section~\ref{sec:discussion}, we discuss the benefits and drawbacks of deploying honeypots across different networks when measuring attacker activity.

\subsection{Discussion and Summary}

Attackers reduce their scanning search space by tailoring their scans towards specific networks and geographic regions. Services hosted in the cloud, especially on SSH-assigned ports, are most likely to be scanned or attacked by a scanner that avoids telescope networks. When filtering for geographic regions, scanners and attackers are most likely to discriminate services hosted in the Asia Pacific---either completely avoiding them or only targeting them. Researchers should be wary of data from only network telescopes, but can use cloud resources to better understand real-world attacks.

\vspace{3pt}
\noindent
\textbf{Temporal consistency.}\quad
When repeating our experiments in July 2020 and July 2022 (Appendix~\ref{app:sub:sec:temporal_net}), scanners targeting services hosted in cloud and education networks continue to significantly avoid telescope networks. 
As in 2021, scanners exhibit less significant preferences when differentiating between cloud and education networks than between different cloud networks. 
Geographic preferences also remain similar (Appendix~\ref{app:sub:sec:temporal_geo}): scanners and attackers are most likely to discriminate services hosted in the Asia Pacific. 
The only different pattern that we see in 2020 is that scanners and attackers targeting SSH/22 are more likely to discriminate amongst geographic regions within the US and EU in 2020 compared to 2021. Nevertheless, discrimination between SSH/22 services hosted in the US and EU is weaker than within the Asia Pacific.
We describe in more detail the similarities and differences in temporal patterns in Appendix~\ref{app:sub:sec:temporal_net} and \ref{app:sub:sec:temporal_geo}.


 \section{Targeted Ports and Protocols}
\label{sub:sec:port_impact}

Building upon our investigation of how attackers source targets, we investigate what protocols they target after having identified an open port.
Researchers recently discovered that the majority of services live on unassigned ports, especially in cloud networks~\cite{lzr}.
In this section, we show that attackers target a different set of protocols than what operators and researchers monitor and analyze. 
Attackers target unexpected protocols (e.g., TLS and Telnet) on IANA-assigned ports (e.g., port~80).
The targeting of unexpected services, which prior work has also found are often more vulnerably configured~\cite{lzr}, causes popular honeypot frameworks and telescopes monitoring HTTP to miss at least 15\% of scanning traffic because they are not engineered to capture unexpected protocol handshakes.  
\looseness=-1

\vspace{3pt}
\noindent
\textbf{Methodology.}\quad
We analyze the traffic destined towards our three /26 networks of honeypots located in the Google, AWS, and \stanford networks,\footnote{To increase our sample size, and since Section~\ref{sub:sec:net} shows that nearly the same set of attackers target both education and cloud networks, we combine data from both education and cloud networks.} all of which are in the same geographic region. We omit the \gn honeypots as they only collect assigned protocol payloads destined towards ports 22, 2222, 23, and 2323 (Section~\ref{sec:methodology}).
Since our non-\gn honeypots do not speak any protocols, our study is limited to only client-first protocols (i.e., only HTTP) to guarantee that a client sends the intended payload immediately after the TCP handshake.
Thus, our results serve as a lower bound, since we are unable to capture unexpected data from a scanner who is waiting for our honeypot to speak a server-first protocol.

We use the open-source scanner LZR~\cite{lzr} to fingerprint unexpected services for 13 of the most popular TCP scanning protocols: HTTP, TLS, SSH, TELNET, SMB, RTSP, SIP, NTP, RDP, ADB, FOX, REDIS and SQL.
We use the GreyNoise API~\cite{greynoise_viz} to label benign and malicious scanning actors. The API labels actors as malicious if the scanning IP was seen actively exploiting services, and benign if the owners of the scanning IPs have undergone a rigorous vetting process~\cite{greynoise_class}. For scanners that GreyNoise does not see or label, we consider the reputation as unknown. We report our results in Table~\ref{table:unex_serv}.

\begin{table}[t]
\centering
\footnotesize
\begin{tabular}{lr|rr}
\toprule

Protocol/Port  & Breakdown & \% Benign & \% Malicious \\ 
\midrule

HTTP/80 & 85\%  & 42\% & 55\% \\
$\sim$HTTP/80 & 15\%  & 42\% & 51\% \\
\midrule
HTTP/8080 & 84\%  & 22\% & 77\% \\
$\sim$HTTP/8080 & 16\%  & 35\% & 49\% \\
\bottomrule
\end{tabular}
\vspace{3pt}
\caption{
\textbf{Scanner-targeted protocols}---%
\textnormal{Malicious scanners target unexpected/unassigned protocols across ports. We define $\sim$Protocol-A/XX to be all protocols that are not Protocol-A\@ that target port XX. Note, the \% of benign and malicious scanners may not sum to 100\% due to a fraction of scanners having unknown status.}
}
\vspace{-10pt}
\label{table:unex_serv}
\end{table}

\vspace{3pt}
\noindent
\textbf{Scanners and attackers target unexpected protocols.}\quad
At least 15\% of scanners that target ports~80 and~8080 do not target the HTTP protocol. 
Rather, 7\% of scanners target TLS, Telnet (0.5\%), SQL (0.4\%), RTSP (0.3\%), SMB (0.3\%), etc. 
Both scanners and attackers target unexpected protocols. 
Across HTTP-assigned ports~80 and~8080, no matter the protocol targeted, at least half of scanners are malicious.
Malicious attackers constitute the majority of scanners that target non-TLS alternative protocols (i.e., Telnet, SMB, etc). 
Scanners from Censys~\cite{censys-2015} are the leading benign organization to find unexpected services. Scanners from various ASes geolocated in China (e.g., ASN\,4134, ASN\,9808) are the leading malicious scanners responsible for exploring unexpected services.

\vspace{3pt}
\noindent
\textbf{Attackers targeting unexpected protocols bypass honeypots and telescopes.}\quad
Popular honeypot frameworks such as Cowrie~\cite{cowrie}, T-Pot~\cite{tpot}, and Kippo~\cite{kippo} by default only perform protocol assigned handshakes on protocol assigned ports. 
Telescopes that do not collect payloads rely on the destination port to derive the target protocol. However, by only performing the assigned handshake or relying on the destination port to fingerprint the protocol, honeypots and telescopes miss at least 15\% of incoming traffic on ports~80 and~8080. 
When possible, honeypots should collect all handshakes across all ports to prevent the underestimation of attacker traffic. 

\section{Limitations and Future Work}
\label{sec:limitations}

Our vantage points provide an IPv4 server’s perspective on scanner behavior, which has several limitations that serve as foundation for future work:

\vspace{3pt}
\noindent
\textbf{Firewalls.}\quad
While none of our honeypots have firewalls, it is possible that a network could transparently drop malicious traffic before they reach our honeypots~\cite{wan20origins}. To mitigate confounding factors, we validate observed patterns across multiple independently-operated networks or geographic regions, which are targeted by tens of thousands of unique IPs and thousands of unique ASes (Table 1). Additionally, we use statistical tests, described in Section 3.3, to report on the statistical significance of the observed patterns.
Future work should measure the prevalence and impact of firewalls across networks. 

\vspace{3pt}
\noindent
\textbf{Honeypot Fingerprinting.} \quad
Scanners occasionally fingerprint honeypots to avoid detection. 
However, the majority of honeypot-fingerprinting requires a scanner to log into the system~\cite{cowrie_issues}, which Gamma honeypots prohibit. A prior exploit that fingerprints Cowrie without logging-in~\cite{bitterharvest} was patched before our data collection.
Nevertheless, other fingerprinting techniques could bias results against sophisticated attackers.
Future work should investigate the prevalence of honeypot fingerprinting across the cloud. 

\vspace{3pt}
\noindent
\textbf{IPv6.} \quad
Unfortunately, we could not study IPv6 scanning patterns, as neither Gamma nor Omega collect/provide IPv6 traffic. Future work should analyze IPv6 scanning patterns in the cloud, since the sparse search space of IPv6~\cite{richter2019scanning} address space will likely surface different scanning patterns. 

\vspace{3pt}
\noindent
\textbf{Protocol Diversity.} \quad
Our analysis focuses on scanning campaigns that target popular protocols over TCP on the cloud.
Scanning campaigns that target unpopular TCP protocols (e.g., SMB, RDP), UDP protocols (e.g., DNS, SNMP) or specialized cloud services (e.g., cloud storage) may target different vulnerabilities and use specialized scanning tools with unique scanning patterns~\cite{cable2021stratosphere,izhikevich2022zdns}, which future work should research. 

\vspace{3pt}
\noindent
\textbf{Temporal Validity.} \quad
The scanning patterns our work surface arise from a set of 1-week data collection periods between 2020--2022. 
Future work should analyze scanning patterns across longer data collection periods, as that may surface different scanning campaigns and new temporal patterns.

\section{Recommendations and Discussion}
\label{sec:discussion}
Our results show that scanners---including known malicious actors---are selective when identifying IPs to scan. Unfortunately, many measurement tools that we use today have made assumptions about scanning that may obstruct our understanding of attacker behavior, particularly when trying to understand how attacks target cloud services and other enterprise networks where vulnerable services are most likely to reside. In this section, we discuss methodological considerations for researchers and service operators attempting to understand and protect against malicious Internet scanning. 

\vspace{3pt}
\noindent
\textbf{Collect scan traffic from networks that host services.}\quad
\label{sec:telescope}
While telescopes have been tremendously useful in understanding some types of attacker behavior, they fail to accurately capture cloud-focused attacks for several reasons: (1)~scanners that target services in cloud and education networks frequently avoid telescopes (Section~\ref{sub:sec:net}); and (2)~most telescopes do not collect payloads, which prevents identifying malicious intent (Section~\ref{sub:sec:suricata}) or the targeted protocol (Section~\ref{sub:sec:port_impact}).
Nevertheless, telescopes do provide the benefit of encompassing large portions of the IP address space and, therefore, a significant sample size. 
Some attacker patterns are visible in telescopes but not cloud services.
For example, identifying scanner address structure preferences (Section~\ref{sec:ip_addrRisk}) would not have been possible using a limited amount of cloud honeypots. 
However, researchers must not assume that the scanning activity a telescope sees is representative of the scanning activity that targets cloud services. Instead, researchers should consider deploying honeypots in networks that house real services.
In many cases, when using telescopes, results should be validated with honeypots deployed in networks that house real services.

\vspace{3pt}
\noindent
\textbf{Consider an IP address' service history.}\quad
Researchers and service operators are often faced with the decision of where to deploy services. The bits and service-search-engine presence of an IP address can increase the likelihood of being attacked, particularly for SSH\@. While likely not a tractable solution for operators to base their security based on an IP's history, researchers need to consider how past activity will affect the research results they collect. Researchers can use search engines (e.g., Censys~\cite{censys-2015} and Shodan~\cite{shodan}) to obtain a history of an IP address.

\vspace{3pt}
\noindent
\textbf{Consider that attackers scan unexpected protocols.}\quad
A significant fraction of services run on unassigned ports. Open source tools for finding unexpected services~\cite{gps,lzr} are now available and search engines have already begun to detect protocols on unassigned ports~\cite{censys2}. Operators should not assume that hiding services on unexpected ports prevents attacker discovery, and researchers should configure honeypots to capture attacker traffic on unexpected ports.


\vspace{3pt}
\noindent
\textbf{Account for differences amongst neighboring IPs.}\quad
Researchers who rely on cloud deployments often do not have large slices of IP address space to devote to honeypots. Consequently, researchers may be tempted to only deploy one honeypot per region~\cite{kelly2021comparative,brown2012HoneypotsIT,bove2019}.
However, our results show that researchers must (1) use more than one honeypot when comparing regions to understand the source of differences; (2) use statistical tests when comparing regions. The majority of scanning activity targets only a subset of the IP address space; it is important to highlight which differences are statistically significant across all honeypots. 

\vspace{3pt}
\noindent
\textbf{Deploy honeypots across geographies, network operators, and IP addresses.}\quad
To maximize attacker traffic (e.g., to populate blocklists or understand scanning behavior), researchers should recognize that significant variation exists even amongst neighboring IP addresses. The IP address itself (e.g., its structure, reputation) should be diversified when deploying honeypots.
Across geographic regions, there is more benefit to deploying a honeypot in a unique geographic region in the Asia Pacific compared to within the US or EU. Across networks, there is more benefit to deploying a honeypot in a different network type (i.e., cloud vs.\ educational) than within the same network type (i.e., AWS vs.\ Google).

\vspace{3pt}
\noindent
\textbf{Consider biases when deploying blocklists.}\quad
Companies and operators often share previously seen malicious IP addresses (e.g., blocklists) and payloads (e.g., payload filters) to help others protect their services. 
Sharing blocklists and payload-filters assumes that the same attackers attack services across geographic locations and networks.  
However, our results show that scanners and payloads differ across continents, especially within the Asia Pacific.
We leave to future work comparing the efficacy of blocklists that source information from different regions.

\vspace{3pt}
\noindent
\textbf{Track attacker trends and update methodologies to protect services accordingly.}\quad
As the Internet and attackers continue to evolve, researchers should reassess the approaches they use to understand network attacks. 
While our results show that attacker preferences remain relatively stable across years, behavioral shifts do occur. For example, deploying honeypots in public clouds may one day become obsolete if the majority of services migrate elsewhere. Further, as the research community develops new tools and data sets to study the Internet, researchers and operators should build protections that can withstand the expectation that attackers will use and abuse the same resources.
\looseness=-1
\section{Conclusion}
In this paper, we showed that Internet-scanning behavior targeting the cloud is nuanced; scanners discriminate between specific IP address structures, regions, and networks. Additionally, attackers have altered their behavior in response to new deployment patterns and public resources, by targeting services on non-standard ports and using Internet search engines to uncover vulnerable services. 
Many of our standard measurement techniques, including using telescopes or only collecting assigned handshakes, have caused us to underestimate and potentially mis-characterize scanner and attacker behavior targeting the cloud. Our work illustrates the importance of reevaluating our measurement instruments and assumptions as the Internet ecosystem and attackers continue to evolve.
\looseness=-1

%
%
%
%

\section*{Acknowledgements}
We thank Hans Hanley, Katherine Izhikevich, Tatyana Izhikevich, Kimberly Ruth, Deepak Kumar, Eric Pauley, Patrick McDaniel, members of the Stanford University security and networking groups, our shepherd, Vasileios Giotsas, and the anonymous reviewers for insightful discussion and comments.
We also thank Daniel Grant, Matt Lehman, Andrew Morris, and the entire GreyNoise team for their invaluable data and support. 
This work was supported in part by the National Science Foundation under awards CNS-1823192, CNS-2120400, CNS-1823192, as well as Google Inc., the NSF Graduate Fellowship DGE-1656518, and a Stanford Graduate Fellowship.

{\footnotesize\bibliographystyle{abbrv}
\bibliography{reference}}

\begin{thebibliography}{10}

\bibitem{honeytrap}
Advanced honeypot framework.
\newblock \url{https://github.com/honeytrap/honeytrap}.
\newblock Accessed on 2022-04-29.

\bibitem{alibaba_ec2}
Alibaba cloud.
\newblock \url{https://us.alibabacloud.com}.
\newblock Accessed on 2022-12-01.

\bibitem{aws_ec2}
Aws {EC2}.
\newblock \url{https://aws.amazon.com/ec2/}.
\newblock Accessed on 2022-12-01.

\bibitem{cowrie}
Cowrie.
\newblock \url{https://github.com/GreyNoise-Intelligence/cowrie}.
\newblock Accessed on 2021-12-28.

\bibitem{cowrie_issues}
Cowrie issue 1102.
\newblock \url{https://github.com/cowrie/cowrie/issues/1102}.
\newblock Accessed on 2021-12-28.

\bibitem{google_ec2}
Google compute engine.
\newblock \url{https://cloud.google.com/compute}.
\newblock Accessed on 2022-12-01.

\bibitem{greynoise_viz}
Greynoise visualizer.
\newblock \url{https://viz.greynoise.io}.
\newblock Accessed on 2022-05-06.

\bibitem{kippo}
Kippo.
\newblock \url{https://github.com/desaster/kippo}.
\newblock Accessed on 2022-05-22.

\bibitem{nmap}
Nmap.
\newblock \url{https://nmap.org/docs.html}.
\newblock Accessed on 2022-05-04.

\bibitem{suricataTriggeredRules}
Suricata rules.
\newblock \url{https://pastebin.com/eqGtVvdX}.

\bibitem{suricataTriggeredRulesREADME}
Suricata rules readme.
\newblock \url{https://pastebin.com/EWSQQkBf}.

\bibitem{suricata}
Suricata user guide.
\newblock \url{https://suricata.readthedocs.io/en/suricata-6.0.5/}.
\newblock Accessed on 2022-05-06.

\bibitem{tpot}
T-pot - the all in one multi honeypot platform.
\newblock \url{https://github.com/telekom-security/tpotce}.
\newblock Accessed on 2021-12-01.

\bibitem{e8home}
Trendmicro: {M}irai-like scanning from {C}hina targets {B}razil.
\newblock
  \url{https://securityonline.info/trendmicro-mirai-like-scanning-from-china-targets-brazil/}.
\newblock Accessed on 2022-05-05.

\bibitem{apac}
What's in a name - exploring the term {APAC}.
\newblock
  \url{https://www.forum-expat-management.com/posts/11371-what-s-in-a-name-exploring-the-term-apac},
  2016.
\newblock Accessed on 2022-05-20.

\bibitem{most_pops_search}
Top 9 {I}nternet search engines used by security researchers.
\newblock \url{https://securitytrails.com/blog/hacker-search-engines}, 2022.
\newblock Accessed on 2022-11-07.

\bibitem{greynoise_class}
Understanding {GreyNoise} classifications.
\newblock
  \url{https://docs.greynoise.io/docs/understanding-greynoise-classifications},
  2022.
\newblock Accessed on 2022-05-10.

\bibitem{akiyoshi2018detecting}
R.~Akiyoshi, D.~Kotani, and Y.~Okabe.
\newblock Detecting emerging large-scale vulnerability scanning activities by
  correlating low-interaction honeypots with darknet.
\newblock In {\em Computer Software and Applications Conference (COMPSAC)},
  volume~2. IEEE, 2018.

\bibitem{anand2023aggressive}
A.~Anand, M.~Kallitsis, J.~Sippe, and A.~Dainotti.
\newblock Aggressive internet-wide scanners: Network impact and longitudinal
  characterization.
\newblock {\em arXiv preprint arXiv:2305.07193}, 2023.

\bibitem{mirai}
M.~Antonakakis, T.~April, M.~Bailey, M.~Bernhard, E.~Bursztein, J.~Cochran,
  Z.~Durumeric, et~al.
\newblock Understanding the {M}irai botnet.
\newblock In {\em USENIX Security Symposium}, 2017.

\bibitem{bano2018scanning}
S.~Bano, P.~Richter, M.~Javed, S.~Sundaresan, Z.~Durumeric, S.~J. Murdoch,
  R.~Mortier, and V.~Paxson.
\newblock Scanning the {I}nternet for liveness.
\newblock {\em ACM SIGCOMM Computer Communication Review}, 2018.

\bibitem{blaise2020detection}
A.~Blaise, M.~Bouet, V.~Conan, and S.~Secci.
\newblock Detection of zero-day attacks: An unsupervised port-based approach.
\newblock {\em Computer Networks}, 180, 2020.

\bibitem{bodenheim2014impact}
R.~C. Bodenheim.
\newblock Impact of the {Shodan} computer search engine on internet-facing
  industrial control system devices.
\newblock Technical report, Air Force Institute of Technology Wright-Patterson
  AFB OH Graduate School of Engineering and Management, 2014.

\bibitem{bove2019}
D.~Bove and T.~Müller.
\newblock Investigating characteristics of attacks on public cloud systems.
\newblock In {\em IEEE International Conference on Cyber Security and Cloud
  Computing (CSCloud)/ IEEE International Conference on Edge Computing and
  Scalable Cloud (EdgeCom)}, 2019.

\bibitem{brown2012HoneypotsIT}
S.~Brown, R.~Lam, S.~Prasad, S.~Ramasubramanian, and J.~Slauson.
\newblock Honeypots in the cloud.
\newblock 2012.

\bibitem{cabana2021threat}
O.~Cabana, A.~M. Youssef, M.~Debbabi, B.~Lebel, M.~Kassouf, R.~Atallah, and
  B.~L. Agba.
\newblock Threat intelligence generation using network telescope data for
  industrial control systems.
\newblock {\em IEEE Transactions on Information Forensics and Security}, 16,
  2021.

\bibitem{cable2021stratosphere}
J.~Cable, D.~Gregory, L.~Izhikevich, and Z.~Durumeric.
\newblock Stratosphere: Finding vulnerable cloud storage buckets.
\newblock In {\em Proceedings of the 24th International Symposium on Research
  in Attacks, Intrusions and Defenses}, pages 399--411, 2021.

\bibitem{censysScanningOptOut}
Censys.
\newblock Opt out of scanning.
\newblock
  \url{https://support.censys.io/hc/en-us/articles/360043177092-Opt-Out-of-Scanning}.
\newblock Accessed on 2022-03-14.

\bibitem{chatziadam2014network}
P.~Chatziadam, I.~G. Askoxylakis, and A.~Fragkiadakis.
\newblock A network telescope for early warning intrusion detection.
\newblock In {\em International Conference on Human Aspects of Information
  Security, Privacy, and Trust}. Springer, 2014.

\bibitem{cramer1946contribution}
H.~Cram{\'e}r.
\newblock A contribution to the theory of statistical estimation.
\newblock {\em Scandinavian Actuarial Journal}, 1946(1), 1946.

\bibitem{censys2}
Z.~Durumeric.
\newblock Censys search 2.0 official announcement.
\newblock
  \url{https://support.censys.io/hc/en-us/articles/360060941211-Censys-Search-2-0-Official-Announcement}.

\bibitem{censys-2015}
Z.~Durumeric, D.~Adrian, A.~Mirian, M.~Bailey, and J.~A. Halderman.
\newblock A search engine backed by {I}nternet-wide scanning.
\newblock In {\em CCS}, 2015.

\bibitem{durumeric2014internet}
Z.~Durumeric, M.~Bailey, and J.~A. Halderman.
\newblock An {I}nternet-wide view of {I}nternet-wide scanning.
\newblock In {\em {USENIX} Security Symposium}, 2014.

\bibitem{durumeric2014matter}
Z.~Durumeric, F.~Li, J.~Kasten, J.~Amann, J.~Beekman, M.~Payer, N.~Weaver,
  D.~Adrian, V.~Paxson, M.~Bailey, et~al.
\newblock The matter of heartbleed.
\newblock In {\em ACM {I}nternet Measurement Conference}, 2014.

\bibitem{durumeric2013zmap}
Z.~Durumeric, E.~Wustrow, and J.~A. Halderman.
\newblock {ZMap}: Fast {I}nternet-wide scanning and its security applications.
\newblock In {\em USENIX Security Symposium}, 2013.

\bibitem{francois2008activity}
J.~Francois, O.~Festor, et~al.
\newblock Activity monitoring for large honeynets and network telescopes.
\newblock {\em International Journal on Advances in Systems and Measurements},
  1(1), 2008.

\bibitem{gadhia2015comparative}
F.~Gadhia, J.~Choi, B.~Cho, and J.~Song.
\newblock Comparative analysis of darknet traffic characteristics between
  darknet sensors.
\newblock In {\em International Conference on Advanced Communication Technology
  (ICACT)}. IEEE, 2015.

\bibitem{griffioen2021scan}
H.~Griffioen, K.~Oosthoek, P.~van~der Knaap, and C.~Doerr.
\newblock Scan, test, execute: Adversarial tactics in amplification {DD}o{S}
  attacks.
\newblock In {\em CCS}, 2021.

\bibitem{spoki}
R.~Hiesgen, M.~Nawrocki, A.~King, A.~Dainotti, T.~C. Schmidt, and
  M.~W{\"a}hlisch.
\newblock Spoki: Unveiling a new wave of scanners through a reactive network
  telescope.
\newblock 2022.

\bibitem{log4shellHTTP}
G.~{I}ntelligence.
\newblock Sample {Log4Shell} ({CVE}-2021-44228) payloads observed in the wild
  by {GreyNoise} {I}ntelligence.
\newblock
  \url{https://gist.github.com/nathanqthai/197b6084a05690fdebf96ed34ae84305}.
\newblock Accessed on 2022-03-14.

\bibitem{irwin2013baseline}
B.~Irwin.
\newblock A baseline study of potentially malicious activity across five
  network telescopes.
\newblock In {\em International Conference on Cyber Conflict (CYCON)}. IEEE,
  2013.

\bibitem{irwin2013source}
B.~Irwin.
\newblock A source analysis of the conficker outbreak from a network telescope.
\newblock {\em SAIEE Africa Research Journal}, 104(2), 2013.

\bibitem{irwin2015observed}
B.~Irwin and T.~Nkhumeleni.
\newblock Observed correlations of unsolicited ip traffic across five distinct
  network telescopes.
\newblock {\em Journal of Information Warfare}, 14(3):1--14, 2015.

\bibitem{izhikevich2022zdns}
L.~Izhikevich, G.~Akiwate, B.~Berger, S.~Drakontaidis, A.~Ascheman, P.~Pearce,
  D.~Adrian, and Z.~Durumeric.
\newblock Zdns: a fast dns toolkit for internet measurement.
\newblock In {\em Proceedings of the 22nd ACM Internet Measurement Conference},
  pages 33--43, 2022.

\bibitem{lzr}
L.~Izhikevich, R.~Teixeira, and Z.~Durumeric.
\newblock {LZR}: Identifying unexpected {I}nternet services.
\newblock In {\em USENIX Security Symposium}, 2021.

\bibitem{gps}
L.~Izhikevich, R.~Teixeira, and Z.~Durumeric.
\newblock Predicting {IP}v4 services across all ports.
\newblock In {\em ACM SIGCOMM Conference}, 2022.

\bibitem{suricataRules}
M.~Jonkman.
\newblock What every {IDS} user should do.
\newblock
  \url{https://doc.emergingthreats.net/bin/view/Main/WhatEveryIDSUserShouldDo}.
\newblock Accessed on 2022-05-03.

\bibitem{kelly2021comparative}
C.~Kelly, N.~Pitropakis, A.~Mylonas, S.~McKeown, and W.~J. Buchanan.
\newblock A comparative analysis of honeypots on different cloud platforms.
\newblock {\em Sensors}, 21(7), 2021.

\bibitem{kramer2015amppot}
L.~Kr{\"a}mer, J.~Krupp, D.~Makita, T.~Nishizoe, T.~Koide, K.~Yoshioka, and
  C.~Rossow.
\newblock Amppot: Monitoring and defending against amplification {DD}o{S}
  attacks.
\newblock In {\em International Symposium on Recent Advances in Intrusion
  Detection}. Springer, 2015.

\bibitem{lagraa2017knowledge}
S.~Lagraa and J.~Fran{\c{c}}ois.
\newblock Knowledge discovery of port scans from darknet.
\newblock In {\em IFIP/IEEE Symposium on Integrated Network and Service
  Management (IM)}. IEEE, 2017.

\bibitem{le2009mitibox}
E.~Le~Mal{\'e}cot.
\newblock Mitibox: camouflage and deception for network scan mitigation.
\newblock In {\em USENIX Workshop on Hot Topics in Security (HotSec)}, 2009.

\bibitem{li2016you}
F.~Li, Z.~Durumeric, J.~Czyz, M.~Karami, M.~Bailey, D.~McCoy, S.~Savage, and
  V.~Paxson.
\newblock You've got vulnerability: Exploring effective vulnerability
  notifications.
\newblock In {\em {USENIX} Security Symposium}, 2016.

\bibitem{li2020survey}
R.~Li, M.~Shen, H.~Yu, C.~Li, P.~Duan, and L.~Zhu.
\newblock A survey on cyberspace search engines.
\newblock In {\em Cyber Security: 17th China Annual Conference, CNCERT 2020,
  Beijing, China, August 12, 2020, Revised Selected Papers 17}, pages 206--214.
  Springer Singapore, 2020.

\bibitem{malecot2013carna}
E.~L. Mal{\'e}cot and D.~Inoue.
\newblock The carna botnet through the lens of a network telescope.
\newblock In {\em International Symposium on Foundations and Practice of
  Security}. Springer, 2013.

\bibitem{moore2002network}
D.~Moore.
\newblock Network telescopes: Observing small or distant security events.
\newblock In {\em USENIX Security Symposium}, 2002.

\bibitem{2004-moore-t20}
D.~Moore, C.~Shannon, G.~Voelker, and S.~Savage.
\newblock Network telescopes: Technical report.
\newblock Technical report, Cooperative Association for Internet Data Analysis
  (CAIDA), 2004.

\bibitem{moura2014bad}
G.~C. Moura, R.~Sadre, and A.~Pras.
\newblock Bad neighborhoods on the internet.
\newblock {\em IEEE communications magazine}, 52(7):132--139, 2014.

\bibitem{nawrocki2021far}
M.~Nawrocki, M.~Jonker, T.~C. Schmidt, and M.~W{\"a}hlisch.
\newblock The far side of {DNS} amplification: tracing the {DD}o{S} attack
  ecosystem from the {I}nternet core.
\newblock In {\em ACM Internet Measurement Conference}, 2021.

\bibitem{nishijima2021verification}
K.~Nishijima, T.~Kondo, T.~Hosokawa, T.~Shigemoto, N.~Kawaguchi, H.~Hasegawa,
  H.~Honda, Y.~Suzuki, T.~Kaji, and O.~Nakamura.
\newblock Verification of the effectiveness to monitor darknet across multiple
  organizations.
\newblock In {\em International Symposium on Computing and Networking Workshops
  (CANDARW)}. IEEE, 2021.

\bibitem{tsunamiBot}
P.~Paganini.
\newblock Multi-vector miner+tsunami botnet with {SSH} lateral movement.
\newblock
  \url{https://securityaffairs.co/wordpress/111761/malware/multi-vector-miner-tsunami-botnet.html}.
\newblock Accessed on 2022-03-14.

\bibitem{pang2004characteristics}
R.~Pang, V.~Yegneswaran, P.~Barford, V.~Paxson, and L.~Peterson.
\newblock Characteristics of {I}nternet background radiation.
\newblock In {\em ACM SIGCOMM conference on Internet measurement}, 2004.

\bibitem{pang2017malicious}
S.~Pang, D.~Komosny, L.~Zhu, R.~Zhang, A.~Sarrafzadeh, T.~Ban, and D.~Inoue.
\newblock Malicious events grouping via behavior based darknet traffic flow
  analysis.
\newblock {\em Wireless Personal Communications}, 96(4), 2017.

\bibitem{pearson1900x}
K.~Pearson.
\newblock On the criterion that a given system of deviations from the probable
  in the case of a correlated system of variables is such that it can be
  reasonably supposed to have arisen from random sampling.
\newblock {\em The London, Edinburgh, and Dublin Philosophical Magazine and
  Journal of Science}, 50(302), 1900.

\bibitem{pham2011honeypot}
V.-H. Pham and M.~Dacier.
\newblock Honeypot trace forensics: The observation viewpoint matters.
\newblock {\em Future Generation Computer Systems}, 27(5), 2011.

\bibitem{pouget2005advantages}
F.~Pouget, M.~Dacier, V.~Pham, et~al.
\newblock On the advantages of deploying a large scale distributed honeypot
  platform.
\newblock In {\em the e-crime and computer evidence conference}, 2005.

\bibitem{prajapati2021shedding}
R.~Prajapati, V.~Honavar, D.~Wu, J.~Yen, and M.~Kallitsis.
\newblock Shedding light into the darknet: scanning characterization and
  detection of temporal changes.
\newblock In {\em International Conference on emerging Networking EXperiments
  and Technologies}, 2021.

\bibitem{raftopoulos2015dangerous}
E.~Raftopoulos, E.~Glatz, X.~Dimitropoulos, and A.~Dainotti.
\newblock How dangerous is internet scanning? a measurement study of the
  aftermath of an internet-wide scan.
\newblock In {\em Traffic Monitoring and Analysis: 7th International Workshop,
  TMA 2015, Barcelona, Spain, April 21-24, 2015. Proceedings 7}, pages
  158--172. Springer, 2015.

\bibitem{richter2019scanning}
P.~Richter and A.~Berger.
\newblock Scanning the scanners: Sensing the {I}nternet from a massively
  distributed network telescope.
\newblock In {\em ACM Internet Measurement Conference}, 2019.

\bibitem{shodan}
SHODAN.
\newblock The search engine for {I}nternet-connected devices.
\newblock \url{https://www.shodan.io/}.
\newblock Accessed on 2021-12-01.

\bibitem{singh2004automated}
S.~Singh, C.~Estan, G.~Varghese, and S.~Savage.
\newblock Automated worm fingerprinting.
\newblock In {\em OSDI}, volume~4, 2004.

\bibitem{soro2019darknets}
F.~Soro, I.~Drago, M.~Trevisan, M.~Mellia, J.~Ceron, and J.~J. Santanna.
\newblock Are darknets all the same? on darknet visibility for security
  monitoring.
\newblock In {\em 2019 IEEE International Symposium on Local and Metropolitan
  Area Networks (LANMAN)}. IEEE, 2019.

\bibitem{torabi2020inferring}
S.~Torabi, E.~Bou-Harb, C.~Assi, E.~B. Karbab, A.~Boukhtouta, and M.~Debbabi.
\newblock Inferring and investigating {I}o{T}-generated scanning campaigns
  targeting a large network telescope.
\newblock {\em IEEE Transactions on Dependable and Secure Computing}, 2020.

\bibitem{bitterharvest}
A.~Vetterl and R.~Clayton.
\newblock Bitter harvest: Systematically fingerprinting low- and
  medium-interaction honeypots at {I}nternet scale.
\newblock In {\em {USENIX} Workshop on Offensive Technologies ({WOOT} 18)},
  Baltimore, MD, Aug. 2018. {USENIX} Association.

\bibitem{wan20origins}
G.~Wan, L.~Izhikevich, D.~Adrian, K.~Yoshioka, R.~Holz, C.~Rossow, and
  Z.~Durumeric.
\newblock On the origin of scanning: The impact of location on {I}nternet-wide
  scans.
\newblock In {\em ACM Internet Measurement Conference}, 2020.

\bibitem{wustrow2010internet}
E.~Wustrow, M.~Karir, M.~Bailey, F.~Jahanian, and G.~Huston.
\newblock Internet background radiation revisited.
\newblock In {\em ACM SIGCOMM conference on Internet measurement}, 2010.

\end{thebibliography}
\appendix


\section{Ethics}

The research carried out in our work does not require IRB approval according to our institutions' policies. Our Institution's IRB is only responsible for human subject research. Our research does not fit any of the criteria: it does not involve biospecimens, interactions with individuals, nor identifiable private information. Thus, our work does not qualify for the IRB process at our institution.

Nevertheless, we agree with and support the mission of minimizing harm when deploying honeypots.
As discussed in Section~\ref{sub:sec:data}, to minimize harm when deploying honeypots, we configure the honeypots to not expose services that are historically prone to being abused for amplification attacks (e.g., DNS open resolver).
Furthermore, our honeypots do not respond to UDP messages, ensuring that no UDP-based DDoS amplification attacks occur. 
The honeypots are also configured to be low-interaction, thereby limiting the size of responses and minimizing the chances of arbitrary code execution triggering a harmful zero-day amplification attack. 
In addition, we continually monitor our honeypots (e.g., ensuring that honeypot IP addresses do not appear in our Telescope logs, monitoring login attempts) to ensure that no attacker has gained control of the honeypots. Our work introduces no new vulnerabilities or exploits that attackers can take advantage of.

\section{How Scanners Filter Network Structures}
\label{app:filt_net}

\begin{figure*}[t]
     \centering
     \begin{subfigure}[]{\columnwidth}
         \centering
         \includegraphics[width=\linewidth]{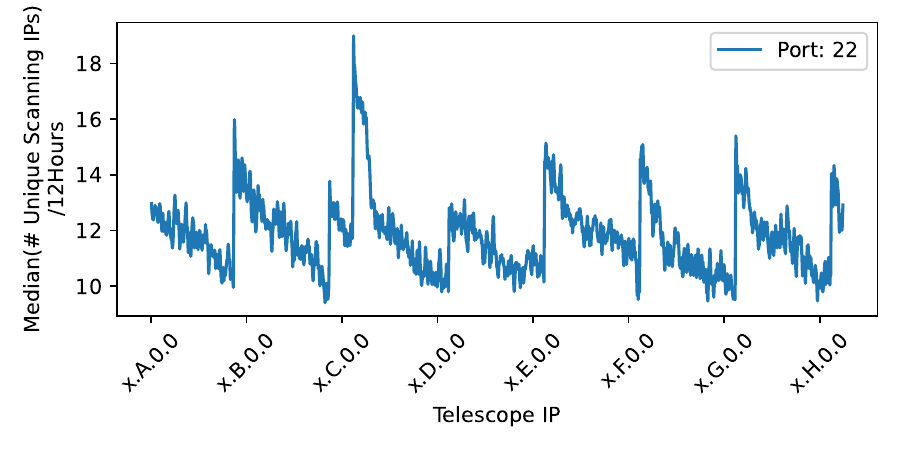}
         \caption{Scanners targeting port~22 are more likely to target the beginning of each /16 network.}
         \label{fig:p22Bias}
     \end{subfigure}
     \hfill
     \begin{subfigure}[]{\columnwidth}
         \centering
         \includegraphics[width=\linewidth]{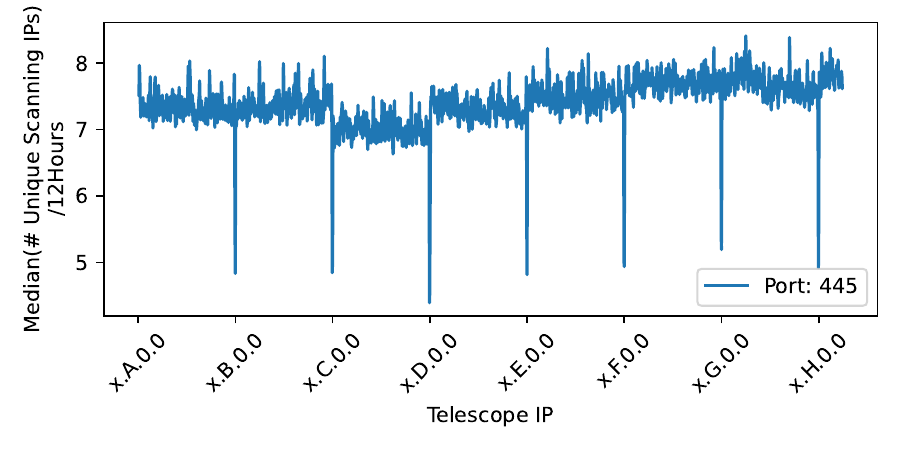}
         \caption{Scanners targeting port~445 are more likely to avoid address with a ``255'' present in any octet.}
         \label{fig:p445Bias}
     \end{subfigure}
     \hfill
     \begin{subfigure}[]{\columnwidth}
         \centering
         \includegraphics[width=\linewidth]{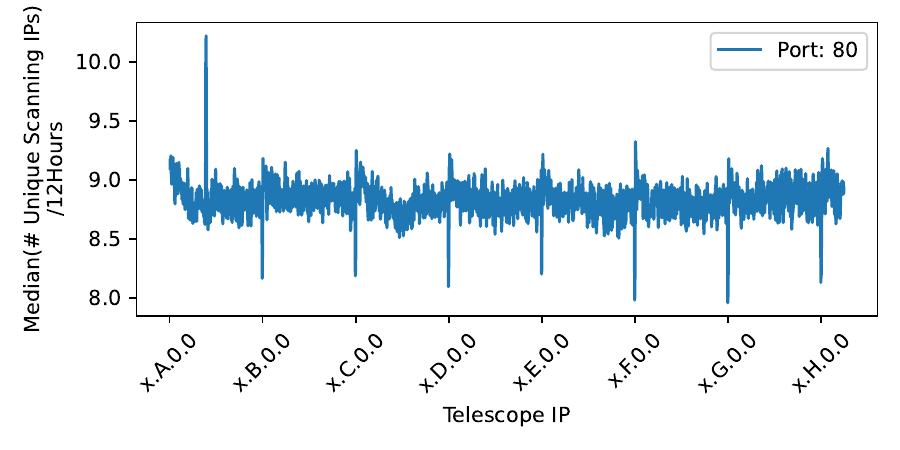}
         \caption{Scanners targeting port~80 are more likely to avoid address with a ``255'' present in any octet.}
         \label{fig:p80Bias}
     \end{subfigure}
     \hfill
     \begin{subfigure}[]{\columnwidth}
         \centering
         \includegraphics[width=\linewidth]{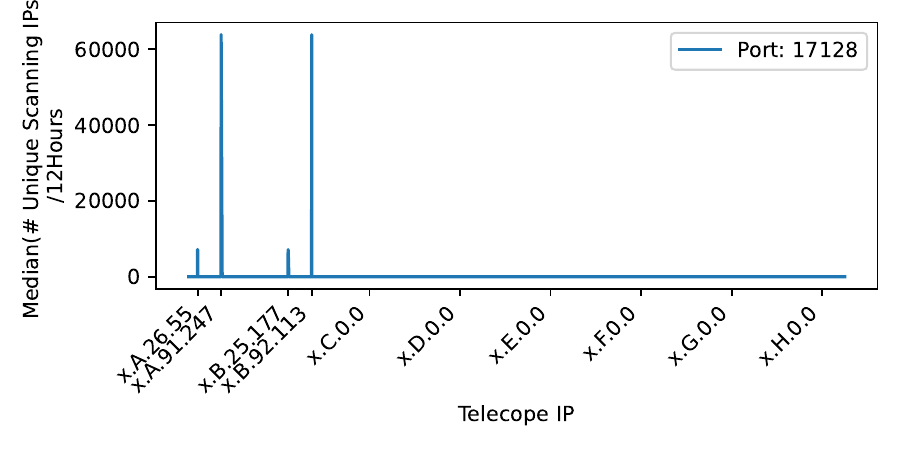}
         \caption{Scanners targeting port~17128 are more likely to target a set of four IP addresses.}
         \label{fig:p17128Bias}
     \end{subfigure}
        \caption{\textbf{Address structure preferences}---%
        \textnormal{Scanners target telescope addresses in a non-uniform manner. To suppress inconsistent outliers, we compute a rolling average of the \# of scanning IPs across every consecutive 512 IPs.}
        }
        \label{fig:bias}
\end{figure*}

Scanners target telescope addresses in a non-uniform manner.
In Figures~\ref{fig:p22Bias}--\ref{fig:p17128Bias}, we compare the number of scanners across neighboring IP addresses in the telescope.
Figures~\ref{fig:p22Bias}--\ref{fig:p80Bias} depict how scanners avoid certain IP address structures, including addresses with a ``255'' present in any octet.  
The avoidance is depicted by the periodical dips in number of unique scanners.
Figure~\ref{fig:p17128Bias} illustrates a single-target preference inside the telescope.

\section{Scanning Patterns Across Time}
\label{app:temporal}

To evaluate the temporal validity of our results, we repeat our experiments from Sections~\ref{sec:sources},~\ref{sec:net_geo}, and~\ref{sub:sec:port_impact} on data collected either exactly one year before (July 1-7, 2020) or after (July 1-7, 2022) the original-experimental data (Section~\ref{sub:sec:meth_temporal}). 
We decide which data to use based upon whether the data comes from the \gn or Honeytrap honeypots; we only have access to \gn deployed honeypots between July 2020--July 2021, and the Honeytrap honeypots were only deployed starting July 2021, and continue to run in July 2022. 

As we will show, attackers and scanners broadly exhibit similar preferences between 2020--2022: they exhibit significant biases when scanning neighboring services, avoid networks without real services, are most likely to scan the Asia Pacific region differently, and scan unexpected protocols on unexpected ports. 
The biggest difference across the years lie in one-off anomalous scanning events, which cause the effect sizes of some patterns to be slightly larger or smaller than in 2021.

\subsection{Discrimination of Neighboring Services}
\label{app:sub:sec:temporal_neighboring}
\begin{table}[t]
\centering
\footnotesize
\resizebox{\columnwidth}{!}{%
\begin{tabular}{lrrrr}
\toprule

&
\multicolumn{2}{c}{SSH/22}
&
\multicolumn{2}{c}{Telnet/23}  \\ \cmidrule(r){2-3}\cmidrule(l){4-5}

Traffic & \% Neighbor- &  Avg. $\phi$ &\% Neighbor-  &   Avg. $\phi$  \\
 
Characteristic & hoods w/ dif & & hoods w/ dif   &  \\
 &  distributions & &distributions &  \\
 &  (n = 53) & & (n = 53) & \\
	\midrule 
Top~3 AS	& 73\% & \cellcolor{red!10}0.23   & 43\% & \cellcolor{red!10}0.38   \\
Fraction Malicious	& 60\% & \cellcolor{blue!10}0.10  & 2\% &  \cellcolor{blue!10}0.14 \\
Top~3 Username	& 74\% & \cellcolor{red!10}0.20 &17\% &\cellcolor{red!10}0.22\\
Top~3 Password	& 19\%& \cellcolor{red!10}0.24 &15\% &\cellcolor{red!10}0.51  \\

\bottomrule \end{tabular}
}
\quad 
\resizebox{\columnwidth}{!}{%
\begin{tabular}{lrrrr}
\toprule

&
\multicolumn{2}{c}{HTTP/80}  
&
\multicolumn{2}{c}{HTTP/All Ports} 
\\ \cmidrule(r){2-3}\cmidrule(l){4-5}

Traffic & \% Neighbor- &  Avg. $\phi$ &\% Neighbor-  &   Avg. $\phi$  \\
 
Characteristic & hoods w/ dif & & hoods w/ dif   &  \\
 &  distributions & &distributions &  \\
 &  (n = 61) & & (n = 61) & \\
	\midrule 
Top~3 AS	& 2\% & \cellcolor{red!10}0.58  & 61\% & \cellcolor{red!10}0.29 \\
Fraction Malicious	& 2\% & \cellcolor{yellow!10}0.21 & 0\% & -    \\
Top~3 Payloads	& 2\% & \cellcolor{red!10}0.50 & 64\% & \cellcolor{red!10}0.54  \\
\bottomrule \end{tabular}
}

	\caption{\textbf{Attackers target neighboring services differently (2020)}---%
	\textnormal{A significantly different set of ASes attack neighboring services with different payloads, including different usernames and passwords. Scanner and attacker behavior is similar to behavior in 2021 (Table~\ref{table:ipaddr_effects}).
	We compare distributions using the chi-square methodology from Section~\ref{sub:sec:chi_meth} and color the effect sizes with its relative magnitude (i.e., blue=``small'', yellow=``medium'', red=``large'').}
	}
\label{table:ipaddr_effects_2020}
\end{table}

In Table~\ref{table:ipaddr_effects_2020}, we show that in July 2020, attackers and scanners still target neighboring services differently. 
When comparing the two years of data, significant differences amongst neighborhoods exist across all points of comparison except the fraction of malicious HTTP requests sent to all ports.
Notably, although significant differences in fraction of malicious traffic did occur in 2021, they all had a very small ($\phi$ < 0.15) effect size.  
We therefore consider the high-level takeaway to be relatively stable across time. 

\subsection{Discrimination Across Networks}
\label{app:sub:sec:temporal_net}

\begin{table*}[t]
\centering
\small
\resizebox{\textwidth}{!}{%
\begin{tabular}{lllllllll}
\toprule

&
\multicolumn{4}{c}{SSH/22}
&
\multicolumn{4}{c}{Telnet/23}  \\ \cmidrule(r){2-5}\cmidrule(l){6-9}

Traffic  & \multicolumn{4}{c}{\# Similar Pairs of Regions in Same Geo-Region/Network} & \multicolumn{4}{c}{\# Similar Pairs of Regions in Same Geo-Region/Network}  \\
\cmidrule(r){2-5}\cmidrule(r){6-9} 
Characteristic & US &  EU  & APAC  &Intercontinental  &  US &  EU  & APAC   & Intercontinental    \\
& (n=31) & (n=19) & (n=40) & (n=267) & (n = 31) & (n=19) & (n=40) &(n=267) \\
	\midrule 
Top~3 AS	& 71\% & 42\% & 30\% & 46\% &94\% &89\% & 77\% & 73\%  \\
Frac Malicious	&  61\%& 63\%& 47\% & 65\% &100\%&100\% & 100\% & 99\%  \\
Top~3 Username	& 55\% & 47\% & 42\%  & 59\% & 100\% & 100\% & 90\% & 87\%  \\
Top~3 Password	& 100\%& 90\%& 97\% &97\% & 100\% & 100\% & 87\% & 87\%   \\

\bottomrule \end{tabular}
}
\quad 
\resizebox{\textwidth}{!}{%
\begin{tabular}{lllllllll}
\toprule

&
\multicolumn{4}{c}{HTTP/80}  
&
\multicolumn{4}{c}{HTTP/All Ports} 
\\ \cmidrule(r){2-5}\cmidrule(l){6-9}
Traffic  & \multicolumn{4}{c}{\# Similar Pairs of Regions in Same Geo-Region/Network} & \multicolumn{4}{c}{\# Similar Pairs of Regions in Same Geo-Region/Network}  \\
\cmidrule(r){2-5}\cmidrule(r){6-9} 
Characteristic & US &  EU  & APAC  &Intercontinental  &  US &  EU  & APAC  &Intercontinental   \\
& (n=31) & (n=19) & (n=40) & (n=267) & (n=31) & (n=19) & (n=40) &(n=267) \\
 
	\midrule 
Top~3 AS	& 100\% &100\% &100\% & 100\% & 34\% & 90\% & 50\% &48\%   \\
Frac Malicious & 100\% &100\% &85\% & 95\% &100\%& 100\%& 100\%&  99\%  \\
Top~3 Payloads	& 100\% &100\% &100\% & 100\% & 53\% & 47\% & 45\% & 54\%  \\

\bottomrule \end{tabular}
}

	\caption{\textbf{Traffic similarities within and between geo-locations (2020)}---%
	\textnormal{Geographic regions within the Asia Pacific regions are much more likely to exhibit statistically significant variation in distribution of different traffic characteristics.}
 }
\label{table:geo_sim_2020}
\end{table*}

\begin{table}[t]
\centering
\scriptsize
\resizebox{\columnwidth}{!}{%
\begin{tabular}{ll|rr|rr|rr}
\toprule
&
&
\multicolumn{2}{c}{Cloud--}
&
\multicolumn{2}{c}{Cloud--} 
&
EDU--\\

&
&
\multicolumn{2}{c}{Cloud}
&
\multicolumn{2}{c}{EDU} 
&
EDU\\
\cmidrule(r){3-4}\cmidrule(l){5-6}\cmidrule(r){7-8}

Traffic & Protocol & \# dif.&  Avg.   &   \# dif. &  Avg.  & \# dif. &  Avg.\\
 & &   Region  & $\phi$   &  Region &   $\phi$  &   Region &   $\phi$  \\
 
 & &  (n=5) & & (n=5) & & (n=1) & \\
	\midrule 
Top~3  & SSH/22 & 2 &  \cellcolor{blue!10} 0.15  & 0 & -& 0 & -  \\ 
AS & TEL/23 & 1&  \cellcolor{blue!10} 0.16  & 3 & \cellcolor{yellow!10}0.30 &1 & \cellcolor{blue!10}0.12 \\
& HTTP/80 & 2  &  \cellcolor{blue!10} 0.20  &4  & \cellcolor{blue!10}0.20 &0 & - \\
& HTTP/All & 4 & \cellcolor{blue!10} 0.24 & 3  & \cellcolor{blue!10}0.15 & 1  &  \cellcolor{blue!10}0.05   \\
\midrule
Top~3 	& SSH/22 & 1 &\cellcolor{blue!10} 0.07  &$\times$&$\times$&$\times$ &$\times$\\
User & TEL/23& 4 & \cellcolor{yellow!10} 0.34 &$\times$&$\times$&$\times$  &$\times$ \\
\midrule
Top~3 & TEL/23& 1 & \cellcolor{blue!10} 0.05  &$\times$&$\times$&$\times$ &$\times$\\
Pwd &  SSH/22 & 2 & \cellcolor{blue!10} 0.11 &$\times$&$\times$&$\times$ &$\times$ \\
\midrule
Top~3 
 & HTTP/80 & 1  & \cellcolor{blue!10}  0.11  & 4&\cellcolor{yellow!10} 0.45 &1 & \cellcolor{yellow!10} 0.34  \\
Payload  & HTTP/All& 4 &\cellcolor{blue!10} 0.24 &3 &\cellcolor{blue!10} 0.16 & 1 &\cellcolor{blue!10} 0.05 \\
\midrule
Frac  & SSH/22 &   1 & \cellcolor{blue!10} 0.02  &$\times$&$\times$&$\times$ &$\times$  \\
Mal & TEL/23& 0 & -  &$\times$&$\times$&$\times$ &$\times$\\
&  HTTP/80 & 0  &- & $\times$ &$\times$ & $\times$ & $\times$\\
& HTTP/All& 0 & -& $\times$& $\times$& $\times$ & $\times$  \\

\bottomrule \end{tabular}
}
\quad 

	\caption{\textbf{Traffic differences across networks: Cloud--Cloud (2020),  Cloud--EDU (2022), and EDU--EDU (2022)}---
 \textnormal{Scanners are are more likely to partially avoid education networks than prefer a specific cloud (e.g., AWS versus Google), similar to 2021 (Table~\ref{table:net_effects}). 
 }
	}
\vspace{-15pt}
\label{table:net_effects_202022}
\end{table}

\begin{table}[t]
\centering
\small
\begin{tabular}{ll|rr|rr}
\toprule
&
&
\multicolumn{2}{c}{Telescope--EDU}
&
\multicolumn{2}{c}{Telescope--Cloud}
\\
\cmidrule(r){3-4}\cmidrule(l){5-6}

Traffic & Protocol & \# dif. &  Avg. $\phi$ & \#  dif. &  Avg. $\phi$ \\
 & &   Region &   &  Region &  \\
 
 & &  (n=2) & &  (n=2)  \\
	\midrule 
Top~3  & SSH/22 &  2  & \cellcolor{red!10}0.57  & 2 & \cellcolor{red!10}0.65  \\ 
AS & TEL/23 & 2& \cellcolor{red!10} 0.54 &2 & \cellcolor{red!10}0.57   \\
& HTTP/80  & 2& \cellcolor{red!10}0.77 & 2 & \cellcolor{red!10}0.78  \\
& Any/All & 2 & \cellcolor{red!10} 0.90  &2  &\cellcolor{red!10} 0.89    \\

\bottomrule \end{tabular}
\quad 

	\caption{\textbf{Different scanners target telescopes (2022)}---
	\textnormal{
    Scanner preferences are even stronger than in 2021 (Table~\ref{table:tel_net_effects}). 
 }
	}
 \vspace{-15pt}
\label{table:tel_net_effects_2022}
\end{table}

In Table~\ref{table:tel_net_effects_2022}, we show that in 2022 scanners that target networks with real services (i.e., clouds and education) are even more likely to originate from ASes that are different than telescope-targeting scanners (e.g., $\phi$ = 0.3 in 2021 vs.\ $\phi$ = 0.89 in 2022).
Similar to 2021, scanners are less likely to differentiate amongst cloud networks and education networks (Table~\ref{table:net_effects_202022}).
Notably, while there are a couple of significant differences in scanning patterns across education networks that were not present in 2021, effect sizes are never large ($\phi$ < 0.34).  
The biggest difference between both years is an anomalous event in which the \merit honeypots get attacked by a set of payloads---bruteforce logins that target router software---that avoid the \stanford honeypots.
Nevertheless, the attack event only causes a medium-significant difference ($\phi$ = 0.34).

\begin{table}[t]
\centering
\footnotesize
\footnotesize
\resizebox{\columnwidth}{!}{%
\begin{tabular}{ll|lr|lr|lr}
\toprule
&
&
\multicolumn{2}{c}{AWS}
&
\multicolumn{2}{c}{Google} 
&
\multicolumn{2}{c}{Linode}

\\ \cmidrule(r){3-4}\cmidrule(l){5-6}\cmidrule(r){7-8}

Traffic & Protocol & Most&  Avg.  &   Most  &  Avg.   & Most .&  Avg.  \\
 & &   Dif.  &   $\phi$   &  Dif. & $\phi$   &   Dif.  & $\phi$    \\

 & &   Region  &     &  Region &   &   Region  &    \\
	\midrule 
Top~3  & SSH/22 &  AP-JP & \cellcolor{yellow!10}0.21 & AP-HK   &\cellcolor{red!10}0.37 & AP-SG &\cellcolor{yellow!10}0.26 \\
AS & TEL/23 & AP-AU &  \cellcolor{red!10}0.27  & AP-KR &   \cellcolor{yellow!10}0.13   & - & -  \\
& HTTP/All & AP-HK & \cellcolor{yellow!10} 0.18 & AP-KR & \cellcolor{yellow!10} 0.26 & - & -  \\
\midrule
Top~3 	& SSH/22 & AP-SG  & \cellcolor{yellow!10}0.20 & AP-HK & \cellcolor{red!10}0.20  & AP-IN & \cellcolor{yellow!10}0.17 \\
User- & TEL/23& CA  & \cellcolor{red!10}0.22 & - & -  & - & - \\
name & & & & & & & \\
\midrule
Top~3  & SSH/22& - & -  & EU-UK & \cellcolor{red!10}0.12  & - & - \\
Pass- & Telnet/23& CA & \cellcolor{red!10}0.20  & -& - & - & -\\
word & & & & & & & \\
\midrule
Top~3 & HTTP/All& AP-JP& \cellcolor{red!10}0.30 & AP-ID & \cellcolor{red!10}0.22 & AP-AU &\cellcolor{blue!10}0.06 \\
Payload & & & & & & & \\
\midrule
Fraction  & SSH/22 & EU-FR & \cellcolor{blue!10}0.11 & EU-NL &  \cellcolor{blue!10}0.12  & AP-IN & \cellcolor{yellow!10}0.28  \\
Malicious & TEL/23& - & - & - & - & - & - \\
& HTTP/80 & - & - &  AP-KR &\cellcolor{red!10} 0.60  & - & -\\
& HTTP/All& US-EAST & \cellcolor{blue!10} 0.10 & - &- & - & - \\

\bottomrule \end{tabular}
}
\quad 

	\caption{\textbf{Geographic traffic patterns (2020)}---
	\textnormal{Asia Pacific regions exhibit the largest statistically significant deviations of traffic distributions compared to other geographic regions within the same network. An ``-'' indicates the absence of statistical significance. 
 }
	}
\label{table:geo_effects_2020}
\end{table}

\begin{table}[t]
\centering
\small
\begin{tabular}{lr}
\toprule

Protocol/Port  & Breakdown \\
\midrule
HTTP/80 & 66\%  \\
$\sim$HTTP/80 & 34\% \\
\midrule
HTTP/8080 & 66\% \\
$\sim$HTTP/8080 & 34\% \\
\bottomrule
\end{tabular}
\caption{
\textbf{Scanners target unexpected/unassigned protocols across ports (2022)}---%
\textnormal{We define $\sim$Protocol-A/XX to be all protocols that are not Protocol-A\@ that target port XX. 
}
}
\label{table:unex_serv_2022}
\end{table}

\subsection{Discrimination Across Regions}
\label{app:sub:sec:temporal_geo}

In Table~\ref{table:geo_sim_2020}, we show that in the year~2020 scanners are most likely to exhibit significant variation when scanning the Asia Pacific region or scanning more than one continent (e.g., US vs.\ EU).
Just like in the year 2021, when comparing individual traffic patterns, regions in the Asia Pacific are most likely to be scanned by an attacker in a significantly-different way (Table~\ref{table:geo_effects_2020}). 
However, we see across both years that anomalous events that affect non-Asia Pacific regions also occur, but are much more rare.
For example, individual scanning attacks targeting SSH/22 within the US and EU cause more significant differences across services within the same continent compared to 2021; nevertheless, differences within the Asia Pacific region remain more stark. 

\subsection{Scanner-Targeted Protocols}

In the year 2022, scanners continue to target unassigned protocols on IANA-assigned ports (Table~\ref{table:unex_serv_2022}).
Scanners are nearly twice as likely to target unassigned targeted protocols in 2022 compared to 2021.
We do not report the breakdown of benign and malicious scanners due to an absence of \gn API data for July 2022.

\end{document}